\documentclass[nofootinbib,aps, amsmath,amssymb,onecolumn,showkeys,letter]{revtex4-2}
\usepackage{xr}

\usepackage{dcolumn}
\usepackage{bm}
\usepackage{comment}
\usepackage{multirow}

\usepackage{tikz, graphics, color, full page, float,epsf,
}

\usepackage{caption}

\usepackage{subfig}

\usepackage{tablefootnote}

\usepackage{pgfplots}
\pgfplotsset{compat=1.9}

\definecolor{forestgreen}{rgb}{0.0,0.5,0.0}

\usepackage{feynmp}
\usepackage{adjustbox}
\usepackage{float}

\usepackage{orcidlink}

\usepackage{calrsfs}
\DeclareMathAlphabet{\pazocal}{OMS}{zplm}{m}{n}

\def\CCC{C$^{3}$~}
\def\CCCfive{C$^{3}$-550}
\def\CCCtwo{C$^{3}$-250}
\def\ee{$e^+e^-$}

\def\CCCnospace{C$^{3}$}
\def\CCCfivenospace{C$^{3}$-550}
\def\CCCtwonospace{C$^{3}$-250}
\def\CCCtwo{C$^{3}$-250 }
\def\CCCfive{C$^{3}$-550 }

\def\GP{GUINEA-PIG }

\begin{document}


\title[Test]{Luminosity and Beam-Induced Background Studies for the \\ Cool Copper Collider}



\author{Dimitrios Ntounis\orcidlink{0009-0008-1063-5620}}  
\email{dntounis@slac.stanford.edu}

\author{Emilio Alessandro Nanni\,\orcidlink{0000-0002-1900-0778}}
\email{nanni@slac.stanford.edu}

\author{Caterina Vernieri\,\orcidlink{0000-0002-0235-1053}}
\email{caterina@slac.stanford.edu}

\affiliation{SLAC National Accelerator Laboratory, 2575 Sand Hill Road, Menlo Park, California 94025, USA  \\ Stanford University, 450 Jane Stanford Way, Stanford, California 94305, USA}

\date[Submitted:]{ Friday 24 May 2024}



\begin{abstract}

A high-energy electron-positron collider has been widely recognized by the particle physics community to be the next crucial step for detailed studies of the Higgs boson and other fundamental particles and processes. Several proposals for such colliders, either linear or circular, are currently under evaluation. Any such collider will be required to reach high lumimosities, in order to collect enough data at a reasonable time scale, while at the same time coping with high rates of background particles produced from beam-beam interactions during the collisions. In this paper, we analyze the luminosity and beam-beam interaction characteristics of the Cool Copper Collider (\CCCnospace) and perform a comparison with other linear collider proposals. We conclude that \CCC can reach the same or higher collision rates as the other proposals, without having to cope with higher beam-induced background fluxes. Thus, \CCC emerges as an attractive option for a future electron-positron collider, benefiting from the collective advancements in beam delivery and final focus system technologies developed by other linear collider initiatives.

\end{abstract}

\maketitle{}
\thispagestyle{plain} 


\section{Introduction}
The Particle Physics community has agreed to pursue a high-energy electron-positron (\ee) collider as the next step beyond the Large Hadron Collider (LHC). It will enable precision measurements of the Higgs boson, the top quark, and electroweak observables, with the potential to uncover signatures of new physics and explore phenomena beyond the Standard Model~\cite{Narain:2022qud, P52023Report, European:2720131}. Several such \ee \ colliders have been proposed over the past few decades, including both linear colliders, such as the Next Linear Collider (NLC)~\cite{NLC_ZDR}, the Compact Linear Collider (CLIC)~\cite{lebrun2012clic}, the International Linear Collider (ILC)~\cite{behnke2013international} and the Cool Copper Collider (\CCCnospace)~\cite{Vernieri_2023}, as well as circular ones, the Future Circular Collider (FCC-ee)~\cite{FCC:2018evy} and the Circular Electron Positron Collider (CEPC)~\cite{cepc_cdr}. For all these machines, one of the main parameters of interest is their instantaneous luminosity $\mathcal{L}_{\mathrm{inst}}$, which determines the rate at which hard scatter (HS)\footnote{A hard scatter event is defined here as an interaction of an electron from one bunch with a positron from the opposite bunch in which a large four-momentum transfer takes place, often resulting in the creation of new particles. } events take place and thus shapes the overall running timeline of each machine.\footnote{The running plan of each machine is roughly determined by the amount of time $T_{\mathrm{run}}$ required to collect a target integrated luminosity $\mathcal{L}_{\mathrm{int}}=\protect\int\limits_{0}^{T_{\mathrm{run}}}{\mathcal{L}_{\mathrm{inst}}(t)\mathrm{d}t}$, necessary for achieving a certain level of precision in the measurement of physical quantities of interest. In this work, we will refer to $\mathcal{L}_{\mathrm{inst}}$ as the instantaneous luminosity or, simply,  luminosity indistinguishably, and we will specifically use ``integrated'' luminosity when referring to $\mathcal{L}_{\mathrm{int}}$.} Unlike the LHC, where beams are \textmu m-sized~\cite{Brüning:782076}, beams focused down to the nm scale are necessary for any \ee \ machine to reach target luminosities of $\mathcal{L}_{\mathrm{inst}} \sim \mathcal{O}(10^{34} \ \mathrm{cm}^{-2} \ \mathrm{s}^{-1})$~\cite{Palmer:1997ku}. At such length scales, 
 the intense electromagnetic (EM) forces between bunch particles give rise to beam-beam interactions that affect the instantaneous luminosity, through self-focusing of the opposing bunches and emission of beamstrahlung (BS) photons, while, at the same time, leading to the creation of beam-induced background (BIB) particles.

Beam-beam interactions constitute an integral component of the design of any future high-energy \ee \ collider and its detectors. They impact the achievable instantaneous luminosity, and thus the runtime required to reach the target integrated luminosity, and dictate the overall running plan and ultimate physics reach of the machine. Beam-beam interactions also contribute to the profile of the beams' luminosity distribution, thus affecting our knowledge of the initial state of the colliding particles, and the level of precision with which the energies and momenta of final-state particles can be constrained. This could in turn limit the sensitivity in extracting physical observables of interest by analyzing collision data. Finally, as the main source of background particles in the interaction region (IR) of the two beams, they dictate the design characteristics of the detectors and the machine-detector interface (MDI), which have to be optimized with the objective of detecting particles from HS events as efficiently as possible while mitigating the contamination from the potentially large numbers of background particles produced.

Therefore, the significance of understanding and evaluating the features of beam-beam effects becomes evident. Extensive simulation studies of these effects have previously been performed, but are mostly limited to evaluating their impact on a specific proposed collider~\cite{NLC_ZDR, Adolphsen:2013kya, Behnke:2013lya, linssen2012physics}. In this work, we utilize such simulations to optimize the beam parameters for \CCC with the objective of maximizing the instantaneous luminosity, without a commensurate increase in the BIB, and additionally perform a comparison between various proposed colliders in terms of the achievable luminosity and the magnitude of the expected BIB. The methodology presented in this note is relevant for any \ee \  collider, and not just for \CCCnospace, which is used here as a test case, since the instantaneous luminosity and BIB profiles have the same dependence on the underlying beam parameters.\footnote{We note that, although the application of this methodology to other linear colliders is rather straightforward, modifications are necessary in the case of circular colliders, to account for the multi-turn nature of the collisions, as opposed to single-pass collisions in linear colliders. Given that this work focuses on the latter, these issues are not addressed here, but are the subject of other recent studies, such as~\cite{Kicsiny:2023pfk}.}
 
This paper is organized as follows: In Sections~\ref{sec:physical_quantities} and~\ref{sec:BIB}, we present an overview of the physical quantities relevant to this study and expand on the concepts of beamstrahlung and beam-induced background. In Section~\ref{sec:opt_c3}, we examine the dependence of the instantaneous luminosity for \CCC on various beam parameters and propose a modified parameter set that leads to luminosity enhancement while keeping the BIB at the same levels. We then proceed to compare various linear collider proposals in terms of their luminosity and BIB characteristics in Section~\ref{sec:comparison}. Finally, we present our conclusions and closing remarks in Section~\ref{sec:conclusions}.


\section{Physical Quantities of interest}
\label{sec:physical_quantities}

The instantaneous luminosity $\mathcal{L}_{\mathrm{inst}}$ of a linear $e^{+}e^{-}$ collider is given by~\cite{schulte_beam_beam}

\begin{equation}
    \mathcal{L}_{\mathrm{inst}}=H_{D}\frac{N_{e}^2 n_{b} f_{r}}{4\pi \sigma_{x}^{*}\sigma_{y}^{*}} = H_{D} \mathcal{L}_{\mathrm{geom}}
    \label{eq:lumi_def}
\end{equation}

\noindent where:

\begin{itemize}
    \item $N_e$ is the number of particles per bunch
    \item $n_b$ is the number of bunches per bunch train
    \item $f_{r}$ is the train repetition rate and
    \item $\sigma_{x,y}^{*}$ are the horizontal and vertical, respectively, root-mean-square (RMS) beam sizes at the Interaction Point (IP).\footnote{From here on, all starred symbols will denote quantities evaluated at the IP.}
\end{itemize}

\noindent The latter are calculated as 

\begin{equation}
    \sigma_{x,y}^{*} = \sqrt{\frac{\epsilon_{x,y}^{*}\beta_{x,y}^{*}}{\gamma}}
    \label{eq:sigmaxy}
\end{equation}

\noindent where $\epsilon_{x,y}^{*},\beta_{x,y}^{*}$ denote, respectively, the normalized emittances and beta functions at the IP in the horizontal ($x$) and vertical ($y$) directions and $\gamma$ is the relativistic Lorentz factor of the beam particles. It is worth noting that $\beta^{*}$ not only determines the beam size at the IP but also how fast it changes around it. Specifically, the inherent betatron oscillations of the beam particles lead to a variation of the beta function with the longitudinal distance $z$ around the IP according to:

\begin{equation}
    \beta_{x,y}(z) = \beta_{x,y}^{*}\left [ 1 + \left (\frac{z}{\beta_{x,y}^{*}} \right)^2 \right ]
\end{equation}

\noindent from which one can see that smaller values of $\beta_{x,y}^{*}$ lead both to a tighter focusing at the IP and a larger increase in the beta function around the IP. This is often referred to as the hourglass effect~\cite{Furman:1991rw, Venturini:2000ji, Zhou:2023ygh}.

Lastly, $H_{D}$ is a dimensionless number, called the enhancement factor, with typical values between 1.5 and 2 for the colliders of interest to this study\footnote{At the Stanford Linear Collider (SLC), operated  between 1989 and 1998 at SLAC, $H_{D}$ was measured to be up to 2.2~\cite{Assmann:2000xv}, providing direct experimental evidence of the importance of this factor for luminosity enhancement.}, which expresses the instantaneous luminosity gain due to the beam-beam interactions taking place at the IP and leading to further ``pinching" of the bunches.  For $H_{D}=1$, that is, when these beam-beam interactions are neglected, the resulting luminosity, calculated using Eq.~\eqref{eq:lumi_def}, is often referred to as the geometric luminosity $\mathcal{L}_{\mathrm{geom}}$. The enhancement factor depends on the so-called disruption parameters, which are defined as the ratio of the bunch length $\sigma_{z}^{*}$ to the focal length $f_{x,y}$ of each beam in the transverse plane due to the attractive forces exerted on it from the opposite beam. For round Gaussian beams, the disruption parameters are given by~\cite{Chen:1987zg, Yokoya:1991qz}

\begin{equation}
    D_{x,y}= \frac{\sigma_{z}^{*}}{f_{x,y}} = \frac{2N_{e}r_{e}\sigma_{z}^{*}}{\gamma\sigma_{x,y}^{*}(\sigma_{x}^{*}+\sigma_{y}^{*})} \quad , \quad \text{where } r_e=\dfrac{1}{4\pi \epsilon_{0}}\dfrac{e^{2}}{m_e c^2} \text{ is the classical electron radius}.
    \label{eq:disruption_param}
\end{equation}

\noindent For linear \ee \ colliders, where flat beams with $\sigma_{x}^{*} \gg \sigma_{y}^{*}$ are used, the disruption parameters have typical values $D_{x}<1, D_{y} \gg 1$, meaning that the vertical motion of the beam particles is significantly disrupted. The enhancement factor additionally depends on the inherent divergence of the beams stemming from the betatron oscillations of the bunch particles which is described with the parameter

\begin{equation}
    A_{x,y} = \frac{\sigma_{z}^{*}}{\beta_{x,y}^{*}}
    \label{eq:A_parameter}
\end{equation}

\noindent with typical values $A_{x} \ll 1$ and $A_{y} \simeq 1$ at \ee \ colliders. Analytical derivations of $H_{D}$ as a function of $D, A$ are notoriously difficult and have only been extracted in a few special cases, most notably for round Gaussian beams with $D \ll 1$~\cite{disruption_round}. For proposed future \ee \ machines with $D_{y} \gg 1$, the dynamics of the beam particles' vertical motion becomes nonlinear and the perturbative analysis used for such calculations is no longer applicable. In such cases, the use of computer simulations for the calculation of $H_{D}$ is necessary.

The two main Particle-In-Cell (PIC) simulation tools that have been used in the High Energy Physics (HEP) community for this purpose are GUINEA-PIG~\cite{Schulte_thesis_GP, Schulte:1999tx} and CAIN~\cite{Chen:1994jt}. These programs rely on the description of the colliding bunches through an ensemble of macroparticles, which are distributed among a three-dimensional grid. At each time step of a simulated bunch crossing (BX), the bunches are split into longitudinal slices, and the Poisson equation is solved for each slice in order to update the electrostatic potential, which in turn governs the evolution of the bunch particles' dynamics.  On top of this classical treatment, the above tools utilize Monte Carlo (MC) methods to simulate the Quantum Electrodynamics (QED) phenomena that arise during beam-beam interactions, such as the emission of Beamstrahlung photons and the production of \ee \ pair particles, that are covered in the next section. Although more modern PIC simulation tools, such as OSIRIS~\cite{osiris} and WarpX~\cite{warpx}, exist that could serve the purpose of beam-beam modeling, they do not currently include all relevant QED processes~\cite{Barklow:2023iav}. For this reason, and in accordance with previous studies performed for ILC and CLIC, we choose to rely on GUINEA-PIG and CAIN simulations for this study. Specifically, although the main simulation results were benchmarked across GUINEA-PIG and CAIN, the results presented in the subsequent sections were solely extracted from GUINEA-PIG simulations. In GUINEA-PIG, the luminosity is computed numerically either by multiplying the charge densities of the intersecting cells and dividing by the cell sizes and time steps, or by adding the contribution from individual  particle collisions. Both methods were found to give virtually identical results in our studies.


\section{Beamstrahlung and Beam-Induced Background}
\label{sec:BIB}

As mentioned earlier, the strong EM beam-beam field in the interaction region due to the intense focusing of the beams leads to energetic synchrotron radiation, called Beamstrahlung. This radiation is characterized by the dimensionless parameter $\Upsilon$, defined as 

\begin{equation}
    \Upsilon = \frac{2}{3} \frac{E_{c}}{E_{0}} = \frac{4}{3} \frac{\hbar \omega_{c}}{\sqrt{s_0}} .
\end{equation}

\noindent  where $\omega_{c}$ is the critical frequency for photon emission and $\sqrt{s_{0}}$ is the nominal (before emission) center-of-mass energy of the colliding particles. The power spectrum of the emitted photons follows the Sokolov-Ternov formula~\cite{sokolov1986radiation}, which can be integrated over to give the average beamstrahlung parameter $\langle \Upsilon \rangle$. For Gaussian beams, this was found to be~\cite{Yokoya:1991qz}

\begin{equation}
    \langle \Upsilon \rangle = \frac{5}{6} \frac{N_{e} r_{e}^{2} \gamma}{\alpha (\sigma_{x}^{*}+\sigma_{y}^{*})\sigma_{z}^{*}} \simeq \frac{5}{6} \frac{N_{e} r_{e}^{2} \gamma}{\alpha \sigma_{x}^{*} \sigma_{z}^{*}}   \ , \text{ where } \alpha \ \text{ is the fine structure constant},
    \label{eq:upsilon}
\end{equation}

\noindent with larger values corresponding to more intense BS photon emission and the approximate equality holding in the limit of flat beams $\sigma_{x}^{*} \gg \sigma_{y}^{*}$. For the proposed future \ee \ colliders, $\langle \Upsilon \rangle$ is typically much less than one. The average beamstrahlung parameter determines the average energy loss $\delta_{E}$ of a beam particle due to beamstrahlung, which is approximately given by~\cite{Yokoya:1991qz}:

\begin{equation}
    \delta_{E} \simeq  \frac{16\sqrt{3}}{5\pi^{3/2}} \frac{r_{e}\alpha N_{e}}{\sigma_{x}^{*}} \langle \Upsilon \rangle \quad \text{for} \  \langle \Upsilon \rangle \ll 1,
    \label{eq:deltaE}
\end{equation}

\noindent as well as the average number and energy fraction of BS photons emitted per beam particle, which can be calculated as: 

\begin{equation}
    n_{\gamma} \simeq  \frac{12}{\pi^{3/2}}\frac{\alpha^2 \sigma_{z}^{*}}{\gamma r_e} \frac{6 \langle \Upsilon \rangle}{5} \quad \text{and} \quad  \left \langle \frac{E_{\gamma}}{E_0}  \right \rangle = \frac{\delta_{E}}{n_{\gamma}} \ .
    \label{eq:n_E_gamma}
\end{equation}

\noindent These BS photons contribute to the creation of BIB particles, most notably additional \ee \ pairs. This occurs through three processes: incoherent pair production, coherent pair production, and trident cascade.

\begin{figure}[H]
    \centering
    \subfloat[Bethe-Heitler]{
        \begin{fmffile}{betheheitler}
        \begin{fmfgraph*}(100,80)
            \fmfleft{i1,i2}
            \fmfright{o1,o2}
            \fmf{photon,label=virtual}{i1,v1}
            \fmf{fermion,label=$e^-$,label.side=right}{v1,o1}
            \fmf{fermion,label=$e^+$,label.side=right}{o2,v1}
            \fmf{photon,label=beamstrahlung}{i2,v1}
            \fmfdot{v1}
        \end{fmfgraph*}
        \end{fmffile}
        \label{fig:betheheitler}
    }\hspace*{-0.1cm} 
    \subfloat[Landau-Lifshitz]{
       \begin{fmffile}{landaulifshitz}
       \begin{fmfgraph*}(100,80)
           \fmfleft{i1,i2}
           \fmfright{o1,o2}
           \fmf{photon,label=virtual}{i1,v1}
           \fmf{fermion,label=$e^-$,label.side=right}{v1,o1}
           \fmf{fermion,label=$e^+$,label.side=right}{o2,v1}
           \fmf{photon,label=virtual}{i2,v1}
           \fmfdot{v1}
       \end{fmfgraph*}
       \end{fmffile}
       \label{fig:landaulifshitz}
    }\hspace*{-0.1cm} 
    \subfloat[Breit-Wheeler]{
        \begin{fmffile}{breitwheeler}
        \begin{fmfgraph*}(100,80)
            \fmfleft{i1,i2}
            \fmfright{o1,o2}
            \fmf{photon,label=beamstrahlung}{i1,v1}
            \fmf{fermion,label=$e^-$,label.side=right}{v1,o1}
            \fmf{fermion,label=$e^+$,label.side=right}{o2,v1}
            \fmf{photon,label=beamstrahlung}{i2,v1}
            \fmfdot{v1}
        \end{fmfgraph*}
        \end{fmffile}
        \label{fig:breitwheeler}
    }
    \caption{Feynman diagrams for the Bethe-Heitler, Landau-Lifshitz, and Breit-Wheeler processes.}
    \label{fig:BS_processes_feynman}
\end{figure}
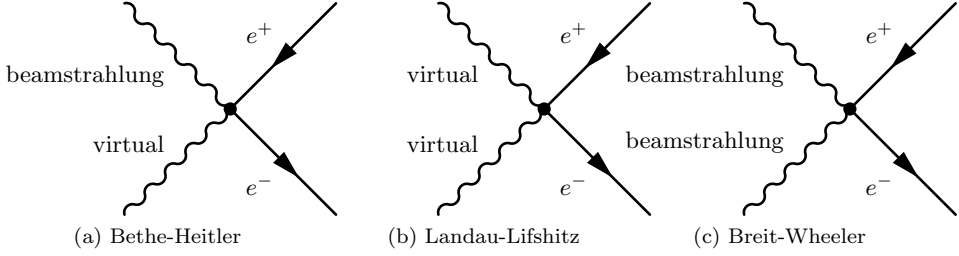

\begin{figure}[H]
    \centering
    \includegraphics[scale=0.6]{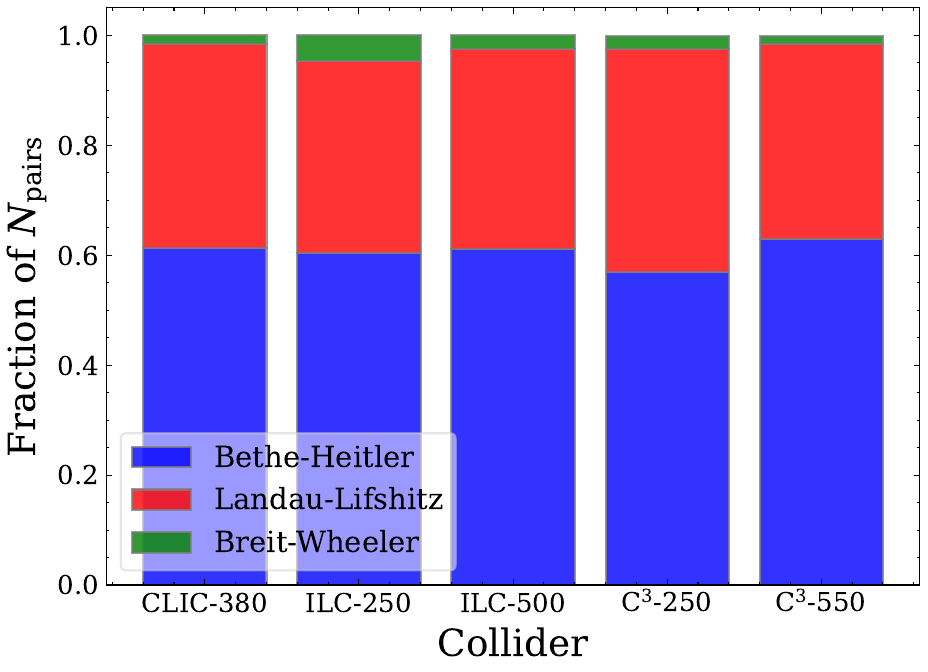}
    \caption{Fraction of incoherently produced  \ee \ pairs from each one of the Bethe-Heitler, Landau-Lifshitz, and Breit-Wheeler processes for various colliders.}
    \label{fig:BS_processes_fraction_Npairs}
\end{figure}

Incoherently produced \ee \ pairs constitute the leading background at any \ee \ collider with  $\langle \Upsilon \rangle \ll 1$  and are created through the interactions of individual photons, either real BS photons or virtual photons that ``accompany" each beam particle. Such pairs are produced through the Bethe-Heitler (BH), Landau-Lifshitz (LL), and Breit-Wheeler (BW) processes, the leading order Feynman diagrams for which are shown in Figure~\ref{fig:BS_processes_feynman}. In the dominant BH process, a beam particle interacts with a real BS photon, whereas in the subdominant LL process, beam particles interact through the exchange of virtual photons. Finally, the BW process is suppressed due to the direct interaction of two BS photons and contributes only at the percent level. Together, these processes result in the creation of order of magnitude $10^{4}-10^{5}$ pairs per bunch crossing for the colliders listed in Table~\ref{tab:beam_params}. A comparison of the relative number of incoherent pairs produced from each process for various colliders is given in Figure~\ref{fig:BS_processes_fraction_Npairs}.

Coherent pair production is the creation of an \ee \ pair through the interaction of a BS photon with the collective EM field of the oncoming beams, instead of with individual particles. This process requires such strong fields that it is exponentially suppressed for  $\langle \Upsilon \rangle \lesssim 0.5$~\cite{Chen:1988ng, Chen:1992ax},  as is the case for all colliders in this study, and leads to a negligible number of \ee \ pairs produced per bunch crossing. Finally, the trident cascade process is the interaction of a virtual photon with the collective EM field of the beams and also results in the production of \ee \ pairs. This process only becomes an important source of background for $\langle \Upsilon \rangle \gg 1$~\cite{Yokoya:1991qz}  and is likewise not relevant in this study.

Beamstrahlung and the resulting beam-induced background have two-fold importance: first, by decreasing the energy of the colliding particles with respect to its nominal value $E_{0}=\sqrt{s_0}/2$, they lead to the widening of their energy distribution, and a so-called luminosity spectrum is created~\cite{chen_diff_lumi}, with contributions to the total luminosity from different center-of-mass energies. The stronger the beam-beam interactions, the more spread out this spectrum is. The fraction of the luminosity in the top $1\%$ of the center-of-mass energy (i.e. for $x \geq 0.99$, with $x=\sqrt{s}/\sqrt{s_0}$ being the center-of-mass energy fraction), denoted as $\mathcal{L}_{0.01}/\mathcal{L}$, is often used to characterize the spread of the luminosity spectrum. 

Second, the produced background particles can reach the detectors of the accompanying experiments and compromise their performance by occupying detector cells with hits from background particles instead of interesting HS events. This can be mitigated, among other approaches, by placing the innermost detector layers further away from the IP or increasing the detector solenoid magnetic field, thus having significant implications for detector design optimization~\cite{Schulte:410057, ILC:2007vrf}.

Lastly, we note that, although our discussion on BIB particles has solely revolved around \ee \ pairs so far, additional species of background particles may be produced through subdominant processes, such as hadron photoproduction~\cite{Schutz:2018ynd}. Despite their smaller production cross-sections, these particles may also affect detector performance and should be taken into account for an accurate estimation of the total BIB at an \ee \ collider.

With the characteristics of beamstrahlung and the beam-induced background analyzed above largely shared among various linear collider proposals, it is useful to juxtapose them. Thus, we perform a comparison of the following proposed linear \ee machines: CLIC,  ILC, and \CCCnospace. Though these projects have been envisaged with various staged approaches, we focus on the sub-TeV runs at center-of-mass energies of 380 GeV for CLIC and at  250, 500 GeV and 250, 550 GeV for ILC and \CCC respectively. Table~\ref{tab:beam_params} shows the main beam parameters of interest for the various colliders\footnote{Some of the parameters have been added for completeness, although they can be readily derived from others, such as $\sigma_{x,y}^{*}$ which can be computed from $\epsilon_{x,y}^{*},\beta_{x,y}^{*}$ using Eq.~\eqref{eq:sigmaxy}, the bunch population $N_e$, which is simply $N_{e}=\frac{Q}{e}$, and the beam power, defined as: 
\begin{equation}
    P_{\mathrm{beam}}=  N_{e} n_{b} f_{r}  \dfrac{\sqrt{s_0}}{2}.
    \label{eq:beam_power}
\end{equation}}. Using these parameters, quantities related to the luminosity and the BIB can be computed, as is shown in Table~\ref{tab:lumi_bkg_params}. Out of them, $\mathcal{L}_{\mathrm{geom}}$, $D_{x,y}$ and $\langle \Upsilon \rangle$ are calculated analytically, using Equations~\eqref{eq:lumi_def},\eqref{eq:disruption_param} and \eqref{eq:upsilon}, respectively, whereas the rest are obtained from GUINEA-PIG simulations. We note that, for all parameters extracted from simulations and for the purpose of a fair evaluation, new simulations were carried out using a common GUINEA-PIG configuration. For CLIC and ILC, our results were benchmarked against the published ones in~\cite{CLIC:2016zwp, Robson:2744946} and ~\cite{ILC_Snowmass}, respectively, and our configuration was tuned in order to ensure reasonable agreement. For \CCCnospace, this marks the first time that such detailed luminosity studies are presented. 

A discussion of the parameters in Tables~\ref{tab:beam_params} and ~\ref{tab:lumi_bkg_params} and their implications is given in Section~\ref{sec:comparison}. Before that, we focus on the beam parameters for \CCCnospace, study their effect on the instantaneous luminosity, and propose a new parameter set, chosen such that luminosity is enhanced while the effect of beamstrahlung and the resulting BIB remains roughly the same.


\begin{table*}[htbp]
    \centering
      \caption{Beam parameters for various linear collider proposals. For \CCCnospace, the baseline beam parameters are given, as found in~\cite{Vernieri_2023}, which we refer to as Parameter Set 1 (PS1) in this work.}
    \label{tab:beam_params}
    \hspace*{-0.9cm}
    \resizebox{1.1\textwidth}{!}{
\begin{tabular}{l c |c|c|c|c|c}
\hline
Parameter & Symbol [unit]  & CLIC~\cite{CLIC:2016zwp} & ILC-250~\cite{ILC_Snowmass} & ILC-500~\cite{ILC_Snowmass} & $\mathrm{C}^3$-250 (PS1)~\cite{Vernieri_2023} & $\mathrm{C}^3$-550 (PS1)~\cite{Vernieri_2023} \\ \hline 
\hline Center-of-mass Energy & $\sqrt{s_0}$ [GeV]  & 380 & 250 & 500 & 250 & 550 \\
RMS bunch length & $\sigma_z^{*}$ [\textmu m]  & 70 & 300 & 300 &  100 & 100 \\
Horizontal beta function at IP & $\beta_x^{*}~[\mathrm{mm}]$  & 8.2 & 13 & 22 & 12 & 12 \\
Vertical beta function at IP & $\beta_y^{*}~[\mathrm{mm}]$ &  0.1 & 0.41 & 0.49 & 0.12 & 0.12 \\
Normalized horizontal emittance at IP & $\epsilon_x^{*}~[\mathrm{nm}]$  & 950 & 5000 & 5000 & 900 & 900 \\
Normalized vertical emittance at IP & $\epsilon_y^{*}~[\mathrm{nm}]$ & 30 & 35 & 35 & 20 & 20 \\
RMS horizontal beam size at IP & $\sigma_x^{*}~[\mathrm{nm}]$  & 145 & 516 & 474 & 210 & 142 \\
RMS vertical beam size at IP & $\sigma_y^{*}~[\mathrm{nm}]$  & 2.8 & 7.7 & 5.9 & 3.1 & 2.1 \\
 Num. Bunches per Train & $n_{b}$  & 352 & 1312 & 1312 & 133 & 75 \\
 Train Rep. Rate & $f_{r}$ $[\mathrm{Hz}]$ & 50 & 5 & 5 & 120 & 120 \\
 Bunch Spacing & $\Delta t_{b}$ [ns]  & 0.5 & 554 & 554 & 5.26 & 3.5 \\
 Bunch Charge & $Q$ [nC]  & 0.83 & 3.2 & 3.2 & 1 & 1 \\
 Bunch Population & $N_{e}$ [$10^{9}$ particles]  & 5.18 & 20.0 & 20.0 & 6.24 & 6.24 \\
 Beam Power & $P_{\mathrm{beam}}$ [MW] & 2.78 & 2.62 & 5.25 & 2.00 & 2.48 \\
 RMS energy spread at the IP & \%  & 0.35 & $\sim$ 0.1 & $\sim$ 0.1 & $\sim$ 0.3 & $\sim$ 0.3 \\
 Crossing Angle\footnote{This is the full angle at which the beams intersect at the IP.} & $\theta$ [rad]  & 0.0165 & 0.014 & 0.014 & 0.014 & 0.014 \\
 Crab Angle\footnote{This is the rotation angle of each beam imparted by the crab cavity, which, when set to half of the crossing angle, can restore head-on collisions.} & $\theta$ [rad] & $0.0165 / 2$ & $0.014 / 2$ & $0.014 / 2$ & $0.014 / 2$ & $0.014 / 2$ \\
 Gradient & [MeV/m] & 72 & 31.5 & 31.5 & 70 & 120 \\
 Effective Gradient & [MeV/m]  & 57 & 21 & 21 & 63 & 108 \\
 Shunt Impedance & [$\mathrm{M}\Omega / \mathrm{m}]$  & 95 &  &  & 300 & 300 \\
 Effective Shunt Impedance &$[\mathrm{M} \Omega / \mathrm{m}]$  & 39 &  & & 300 & 300 \\
 Site Power & $[\mathrm{MW}]$  & 168 & 125 &  173 & $\sim 150$ & $\sim 175$ \\
 Length & $[\mathrm{km}]$  & 11.4 & $20.5$ & $31$ & 8 & 8 \\
 $\mathrm{L}^*$\footnote{$\mathrm{L}^*$ denotes the distance from the final quadrupole magnet (quad) to the IP.} & $[\mathrm{m}]$  & 6 & 4.1 & 4.1 & 4.3 & 4.3 \\
\hline \hline
\end{tabular}
}
\end{table*}

\vspace{1cm}

\begin{table*}[htbp]
\centering
\caption{Luminosity and beam-induced background related quantities for various linear collider proposals. The horizontal line after the fourth row separates the quantities in the ones calculated analytically using Equations~\eqref{eq:lumi_def}, \eqref{eq:disruption_param} and \eqref{eq:upsilon} (top) and the ones whose values are extracted from \GP \space simulations (bottom).}\label{tab:lumi_bkg_params}

\hspace*{-0.9cm}
\resizebox{1.1\textwidth}{!}{
\begin{tabular}{l c|c|c|c|c|c}
\hline
Parameter & Symbol [unit]  & CLIC & ILC-250 & ILC-500 & $\mathrm{C}^3$-250 (PS1) & $\mathrm{C}^3$-550 (PS1)\\ \hline 
\hline 
Geometric Luminosity  & $\mathcal{L}_{\mathrm{geom}}$ $\left[10^{34}  \mathrm{~cm}^{-2} \mathrm{~s}^{-1}\right]$  & 0.91 & 0.53 & 0.74 & 0.75 & 0.93 \\
Horizontal Disruption & $D_x$  & 0.26 & 0.51 & 0.30 & 0.32 & 0.32 \\
Vertical Disruption &  $D_y$  & 13.1 & 34.5 & 24.3 & 21.5 & 21.5 \\
Average Beamstrahlung Parameter & $\langle \Upsilon \rangle$   & 0.17 & 0.028 & 0.062 & 0.065 & 0.21 \\ \hline
Total Luminosity & $\mathcal{L}$ $\left[10^{34}  \mathrm{~cm}^{-2} \mathrm{~s}^{-1}\right]$  & 1.67 & 1.35 &  1.80 & 1.35 & 1.70 \\
Peak luminosity fraction & $\mathcal{L}_{0.01}/\mathcal{L}$ $[\%]$ & 59 & 74 & 64 & 73 & 52 \\ 
Enhancement Factor&  $H_D$  & 1.8 & 2.6 & 2.4 & 1.8 & 1.8 \\
Average Energy loss & $ \delta_{E}$ $[\%]$  & 6.9 & 3.0 & 4.5 & 3.3 & 9.6 \\
Photons per beam particle & $n_{\gamma}$ & 1.5 &  2.1 & 1.9 & 1.4  & 1.9\\
Average Photon Energy fraction & $\langle E_{\gamma}/E_{0} \rangle$ $[\%]$  & 4.6 & 1.4 & 2.3 & 2.5 & 5.1 \\
Number of incoherent particles/BX & $N_{\mathrm{incoh}}$ [$10^{4}$]  & 6.0 & 13.3 & 18.5 & 4.7 & 12.6 \\
Total energy of incoh. particles/BX & $E_{\mathrm{incoh}}$ [TeV]  & 187 & 117 & 439 & 58 & 644 \\  \hline \hline
\end{tabular}
}

\end{table*}


\section{Beam parameter Optimization for \CCC}
\label{sec:opt_c3}

\subsection{Introduction}

The proposed physics program for \CCC is laid out in Ref.~\cite{Vernieri_2023} and consists of a staged approach with runs at center-of-mass energies of 250 and 550 GeV, with the goal of collecting, similar to ILC, integrated luminosities of $2 \ \mathrm{ab}^{-1}$ and $4 \ \mathrm{ab}^{-1}$, respectively, after 10 years of data taking at each energy. In order to achieve this goal, the target luminosities for \CCC have been established to be: 

\begin{align*}
   \mathcal{L}_{\text{C}^3 - 250}^{(\mathrm{target})} &=  1.3 \cdot 10^{34} \ \mathrm{cm}^{-2} \ \mathrm{s}^{-1} \\ 
    \mathcal{L}_{\text{C}^3 - 550}^{(\mathrm{target})} &=  2.4 \cdot 10^{34} \ \mathrm{cm}^{-2} \ \mathrm{s}^{-1}
\end{align*}

\noindent The proposed beam parameters for \CCCnospace-250, as summarized in Table~\ref{tab:beam_params}, are sufficient to reach the corresponding target luminosity. On the other hand, achieving the target luminosity at 550 GeV is more challenging.  While it is generally true for linear colliders that the luminosity increases linearly with $\sqrt{s}$ when keeping all other beam parameters fixed, as can be seen from Equations~\eqref{eq:lumi_def} and \eqref{eq:sigmaxy}, the operation of \CCC at 550 GeV requires an increased gradient of $120 \ \mathrm{MeV}/\mathrm{m}$, which, in turn, necessitates a reduction of the flat top from 700 to 250 ns. A roughly halved number of 75 bunches per train is expected to fit within this shortened flat top, which mitigates the luminosity increase due to the higher center-of-mass energy. For this reason, achieving the target luminosity for \CCCnospace-550 requires further optimization of the beam parameters.

In this section, we begin by optimizing the luminosity for the \CCCnospace-550 stage with respect to the vertical $\epsilon_{y}^{*}$ and horizontal $\epsilon_{x}^{*}$ emittance, the bunch length $\sigma_{z}^{*}$ and the vertical waist shift $w_{y}$, which will be introduced in the following. As a result of this optimization process, we propose a new parameter set for \CCC with which the luminosity goal at 550 GeV is achieved. We then compare this parameter set with the nominal beam parameters (PS1) and evaluate its performance for \CCC operated at 250 GeV. We observe, at both center-of-mass energies, an enhanced luminosity, without at the same time enhancing the BIB. We proceed to investigate the luminosity dependence on additional beam parameters, namely the beta function and the beam offset, which were not directly taken into account in the optimization process but nevertheless influence the luminosity. Finally, we discuss power consumption considerations related to the luminosity requirements for \CCCnospace.  

\subsection{Parameter Optimization for \CCCnospace-550}
\label{subsec:opt_c3-550}

\begin{figure*}[h!]

\subfloat[\centering{\label{fig:c3_550_emittance_scan}}]{%
\hspace*{-0.9cm}
    \includegraphics[scale=0.41]{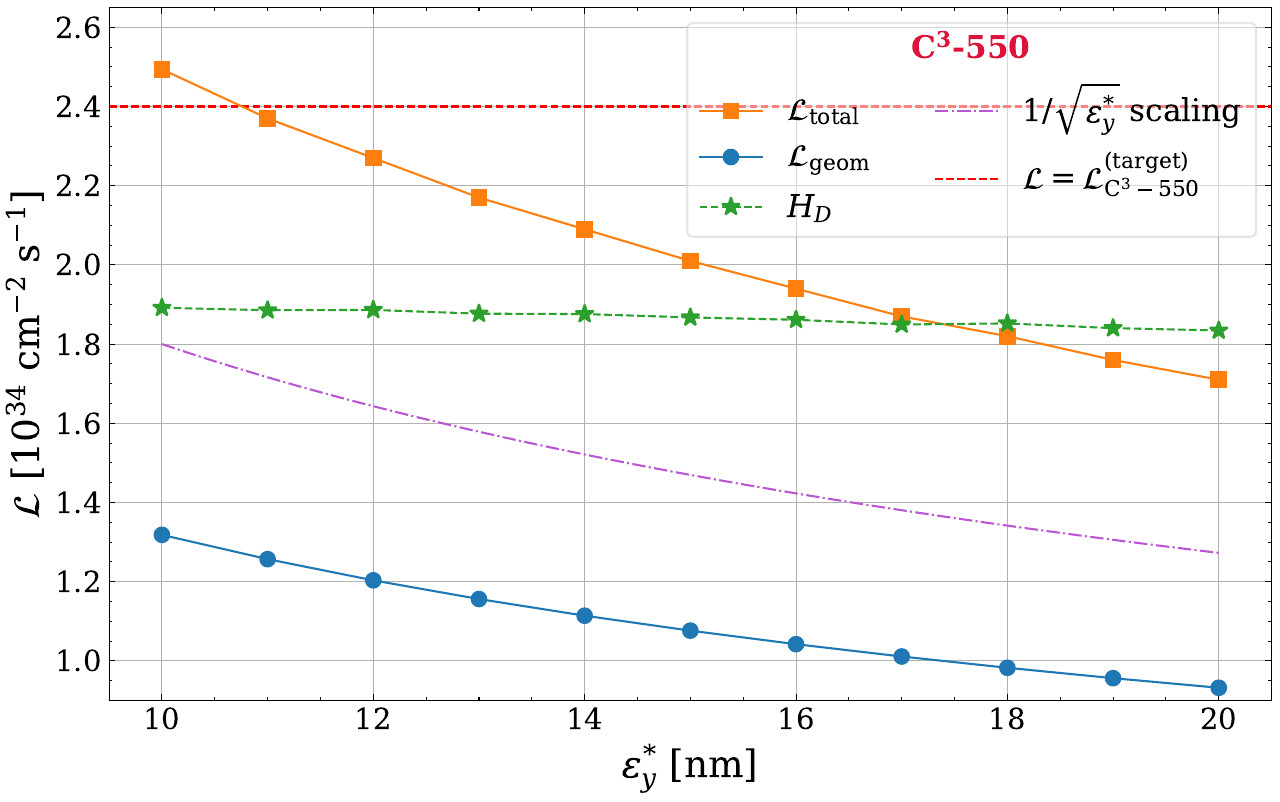}
} 
\subfloat[\centering{\label{fig:c3_550_waist_shift_scan}}]{%
   \includegraphics[scale=0.41]{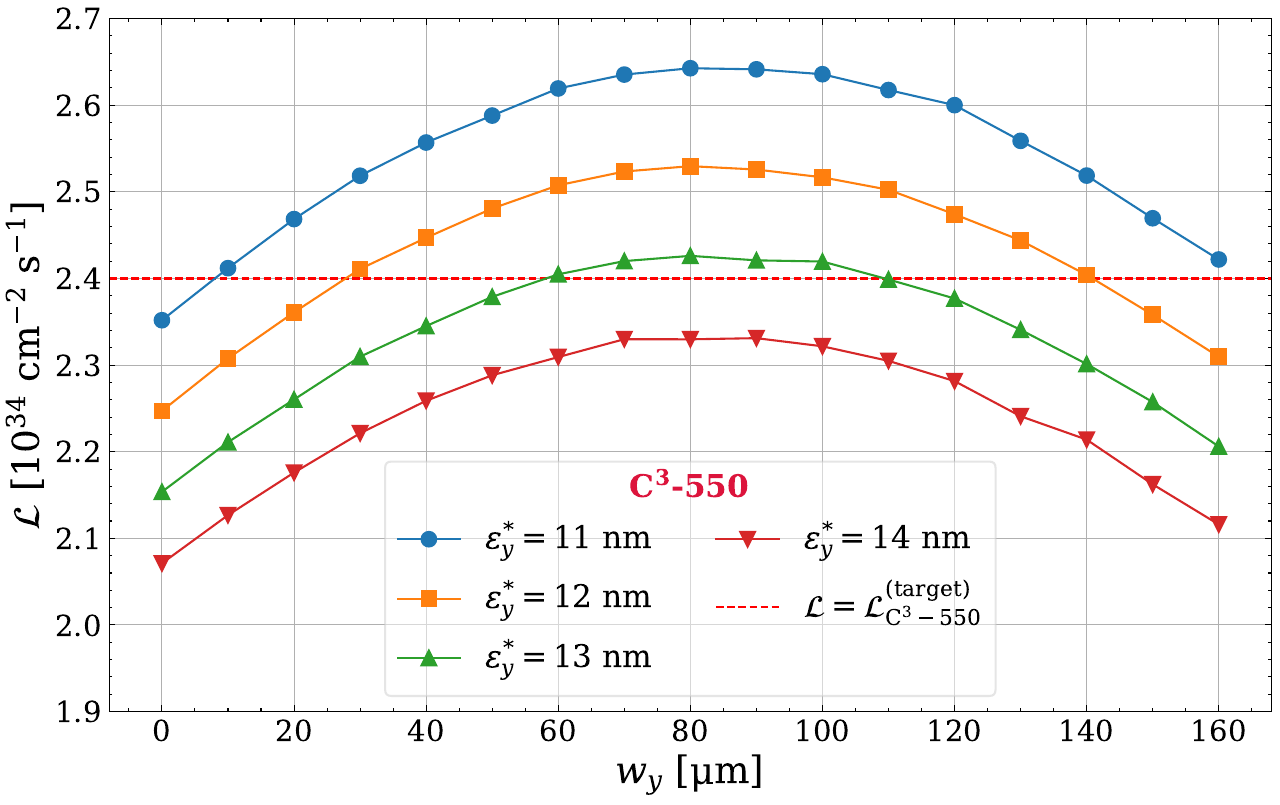}
}

\caption{Instantaneous luminosity for \CCCnospace-550 as a function of (a) the vertical emittance $\epsilon_{y}^{*}$ and (b) the vertical waist shift $w_y$, while keeping all other parameters fixed to their nominal values. In (a), in addition to the total instantaneous luminosity, the geometric luminosity $\mathcal{L}_{\mathrm{geom}}$, the enhancement factor $H_{D}$ and the scaling of the luminosity in the absence of beam-beam interactions $1/\sqrt{\epsilon_{y}^{*}}$ are given.  In both plots, the horizontal red dashed line indicates the target luminosity for \CCCnospace-550.}
\label{fig:C3_550_lumi_scans_epsilony_waist}
\end{figure*}

In order to optimize the beam parameters for \CCCnospace-550, we first note that preliminary simulations of emittance growth in the \CCCnospace-250 main linac and Beam Delivery System (BDS),  assuming 10 nm vertical emittance at injection, indicate that the vertical emittance at the IP $\epsilon_{y}^{*}$ can be maintained below 20 nm~\cite{white2023lcws,tan2024c3}. Anticipating similar emittance growth levels at 550 GeV as well, we perform a scan of the luminosity for \CCCnospace-550 with respect to $\epsilon_{y}^{*}$, whilst keeping all other beam parameters at their nominal values. To obtain this scan, we run separate \GP simulations at each $\epsilon_{y}^{*}$ point. For this, and all subsequent simulations, we have assumed a 3D Gaussian distribution for the initial state of the beams.  The results of the simulations are shown in Figure~\ref{fig:c3_550_emittance_scan} and indicate that both the geometric luminosity and the full luminosity, i.e. when taking into account beam-beam effects, scale as $1/\sqrt{\epsilon_{y}^{*}}$, with the enhancement factor remaining roughly flat, due to the beam-beam interactions being driven by the horizontal beam parameters. As can be seen, achieving the target luminosity without any other modifications in the nominal beam parameters would necessitate a reduction of the vertical emittance down to around 11 nm.

\begin{figure}[h]
\centering

\begin{tikzpicture}
\begin{axis}[
    xlabel={$z/\beta_y^*$},
    ylabel={$\pm \ \sigma_y/\sigma_{y}^{*}$},
    xmin=-2.5, xmax=2.5,
    ymin=-3, ymax=3,
    legend pos=north west,
    ymajorgrids=true,
    grid style=dashed,
]

    \addplot[
        color=blue,
        mark=none,
        smooth,
        thick,
        domain=-3:3,
        samples=500,
    ] {sqrt(1+((x-0.8)/1.2)^2)};

    \addplot[
        color=blue,
        mark=none,
        smooth,
        thick,
        domain=-3:3,
        samples=500,
    ] {-sqrt(1+((x-0.8)/1.2)^2)};

    \addplot[
        color=purple,
        mark=none,
        smooth,
        thick,
        domain=-3:3,
        samples=500,
    ] {sqrt(1+((x+0.8)/1.2)^2)};

    \addplot[
        color=purple,
        mark=none,
        smooth,
        thick,
        domain=-3:3,
        samples=500,
    ] {-sqrt(1+((x+0.8)/1.2)^2)};
\draw[<->, thick] (axis cs:-0.8,0.8) -- (axis cs:-0.01,0.8) node[midway, below] {$w_y$};;
\draw[<->, thick] (axis cs:+0.01,0.8) -- (axis cs:0.8,0.8) node[midway, below] {$w_y$};;
\draw[->, thick,color=purple] (axis cs:-1.5,-0.8) -- (axis cs:-0.7,-0.8) node[midway, above] {beam moving right};;
\draw[<-, thick,color=blue] (axis cs:0.7,-0.8) -- (axis cs:1.5,-0.8) node[midway, above] {beam moving left};;
\draw[dashed,thick,color=forestgreen](axis cs:0,-3) -- (axis cs:0,3)
node[midway] {IP};;
\end{axis}
\end{tikzpicture}
 \caption{Illustration of a positive vertical waist shift $w_{y}$ for two colliding bunches. The horizontal axis shows the longitudinal direction normalized to the vertical beta function at the IP $\beta_{y}^{*}$ and the vertical axis shows the RMS spot size in the y-direction $\sigma_{y}$ normalized to its value at the IP $\sigma_{y}^{*}$. The green dashed line at $z=0$ indicates the position of the IP. For this figure, \CCC beam parameters and a vertical waist shift of $w_{y} = 0.8\sigma_{z}^{*}$ have been assumed.}
\label{fig:waist_shift_explanation}
\end{figure}
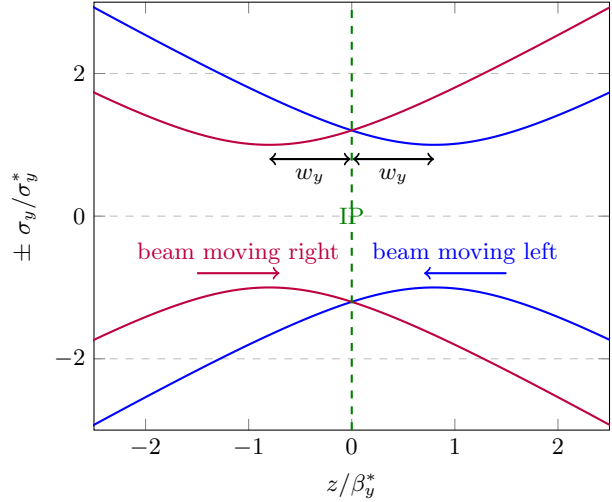

If such vertical emittance values prove challenging for the \CCC accelerator design, the target luminosity could be achieved by introducing a shift $w_{y}$ in the longitudinal position of the vertical waist of the bunches. The concept of vertical waist shift is illustrated in Figure~\ref{fig:waist_shift_explanation} and lies in longitudinally displacing the focal point (waist) of the two opposite bunches in the vertical direction with respect to the IP. When the vertical waists are positioned before the IP, i.e. $w_{y}>0$, beam-beam interactions focus opposite bunches such that the actual waists coincide at the IP, leading to an increase in the instantaneous luminosity. The impact of waist-shifts has previously been investigated in the context of ILC and CLIC~\cite{schulte_beam_beam} and has been shown to lead to a roughly $ 10 \%$ luminosity gain for waist shift values similar to the bunch length: $w_{y} \sim \sigma_{z}^{*}$. In Figure~\ref{fig:c3_550_waist_shift_scan}, the instantaneous luminosity as a function of the vertical waist shift is shown for different values of vertical emittance. We notice that, regardless of the vertical emittance value, the luminosity is maximized at waist shifts of around $80$  \textmu m, which corresponds to $0.8\sigma_{z}^{*}$. For such waist shifts, the target luminosity for \CCCnospace-550 can be achieved for larger vertical emittances up to about 13 nm.

\begin{figure*}[h!]
\subfloat[\centering\label{fig:c3_550_bunch_length}]{%
    \hspace*{-0.9cm}
    \includegraphics[scale=0.41]{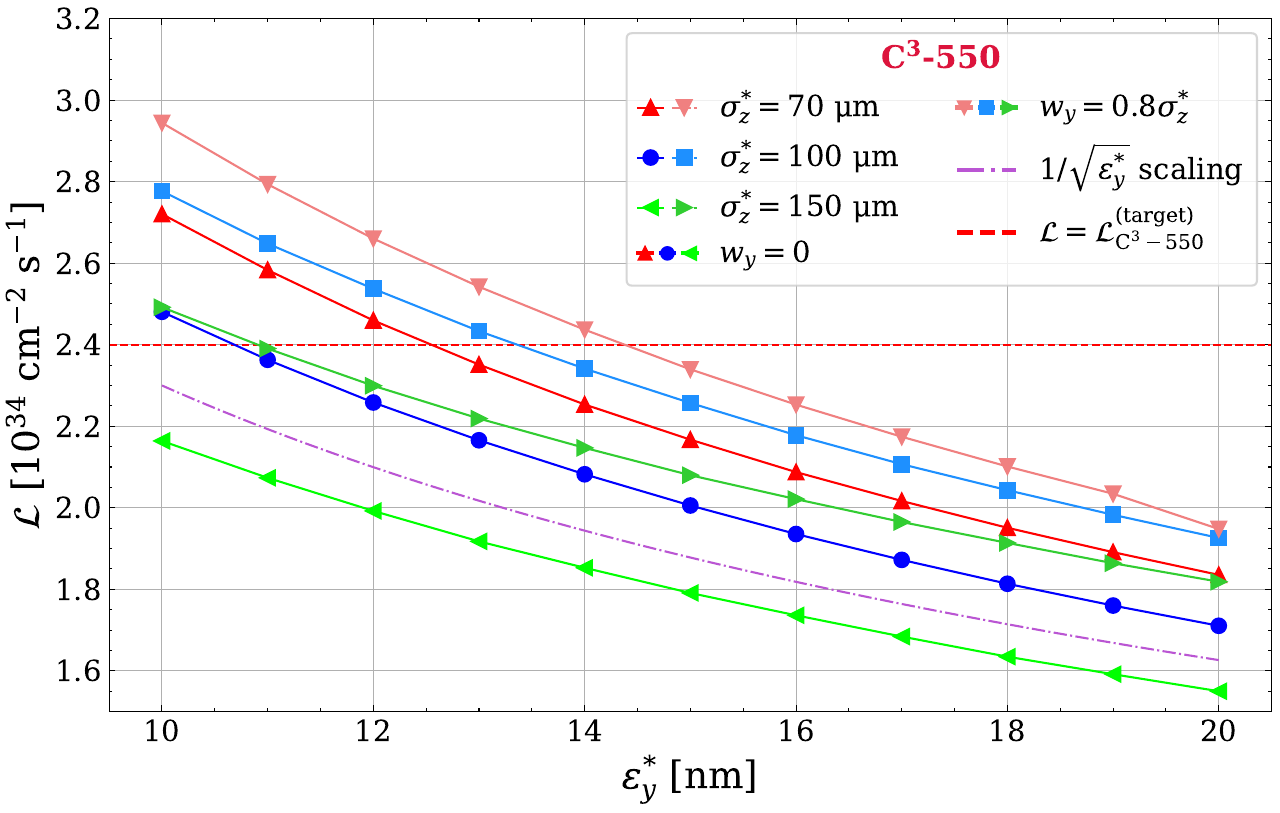}
}
\subfloat[\centering\label{fig:c3_550_horizontal_emittance}]{%
    \includegraphics[scale=0.41]{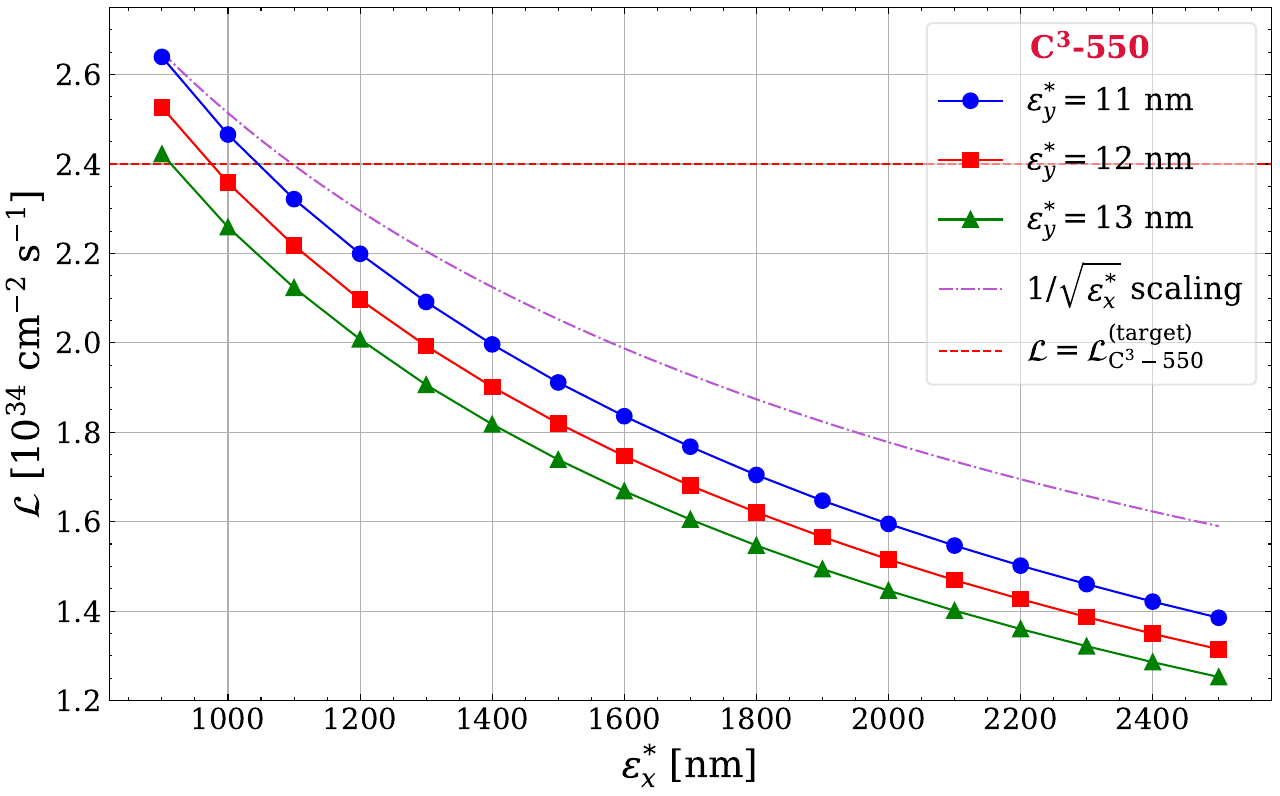}
}
\caption{Instantaneous luminosity for \CCCnospace-550 as a function of (a) the vertical $\epsilon_{y}^{*}$ and (b) the horizontal $\epsilon_{x}^{*}$ emittance and further modified beam parameters as shown in each figure. All other beam parameters are kept to their nominal values. The scaling of the luminosity in the absence of beam-beam interactions $1/\sqrt{\epsilon^{*}}$ is also given in each case. In both plots, the horizontal red dashed line indicates the target luminosity for \CCCnospace-550.}
\label{fig:C3_550_lumi_scans_sigmaz_epsilonx}
\end{figure*}

Additional optimization of the \CCCnospace-550 luminosity can be achieved by modifying the bunch length $\sigma_{z}^{*}$. Although this parameter does not directly affect the geometric luminosity, Eq.~\eqref{eq:lumi_def}, it does have an impact on the beam-beam interactions, Eq.~\eqref{eq:upsilon}, and can thus modify the enhancement factor $H_{D}$. The bunch length for \CCCnospace, taking into account bunch compression limitations in the current \CCC accelerator concept, is foreseen to have a minimum (maximum) allowed value of 70 (150) \textmu m.\footnote{ In general, smaller bunch lengths would lead to stronger longitudinal wakefields,  thus increasing the energy spread of the beam particles, whereas larger bunch lengths would enhance beam tilt induced by transverse wakefields, leading to emittance growth. For \CCCnospace, a bunch length of 100 \textmu m was found to be a reasonable compromise between the two effects.} For these extreme values, as well as the nominal bunch length of 100 \textmu m, we estimate the luminosity for \CCCnospace-550 assuming various values of the vertical emittance $\epsilon_{y}^{*}$ and vertical waist shifts $w_{y}$ of 0 and $0.8\sigma_{z}^{*}$. The results are shown in Figure~\ref{fig:c3_550_bunch_length}. As expected, smaller $\sigma_{z}^{*}$ values lead to stronger beam-beam interactions and, thus, larger enhancement factors. We also note that parameter configurations with waist shifts consistently achieve higher instantaneous luminosities.

For a bunch length of $\sigma_{z}^{*}=70 $ \textmu m and a waist shift of $w_{y}=0.8\sigma_{z}^{*}$, the target luminosity can be achieved for vertical emittances up to 14 nm. However, such small bunch length values come at the cost of increased beam-beam interactions, with a potentially significant impact on detector performance and the luminosity spectrum. Indeed, a significant broadening of the luminosity spectrum is observed when reducing $\sigma_{z}^{*}$ to 70 \textmu m~\cite{supplemental}. Owing to this, as well as the fact that, according to Figure~\ref{fig:c3_550_bunch_length}, luminosity gains are only minimal when reducing  $\sigma_{z}^{*}$ from 100 to 70 \textmu m, we choose to keep the bunch length at 100 \textmu m.

Finally, to reduce beam-beam interactions while achieving the target luminosity, we investigate the effect of varying the horizontal emittance $\epsilon_{x}^{*}$, which affects the beamstrahlung parameter according to Eq.~\eqref{eq:upsilon}. We perform luminosity scans for $\sigma_{z}^{*} = 100$ \textmu m, $w_{y}=0.8\sigma_{z}^{*}$ and $\epsilon_{y}^{*}=11,12$ and 13 nm, i.e. three vertical emittance values for which luminosities close to the target one can be achieved according to Figure~\ref{fig:C3_550_lumi_scans_epsilony_waist}. The luminosity scans are shown in Figure~\ref{fig:c3_550_horizontal_emittance}. Since increasing $\epsilon_{x}^{*}$ reduces beam-beam interactions and thus the enhancement factor $H_{D}$, the luminosity drops faster than the expected $1/\sqrt{\epsilon_{x}^{*}}$ dependence of the $\mathcal{L}_{\mathrm{geom}}$ term in Eq.~\eqref{eq:lumi_def}. Due to this, the horizontal emittance cannot be adjusted to values significantly above the PS1 value of 900 nm without considerable luminosity losses. We found that a moderate increase of the horizontal emittance from 900 to $1000 \ \mathrm{nm}$ can still achieve the target luminosity for $\epsilon_{y}^{*}= 12 \ \mathrm{nm}$, as can be seen in Figure~\ref{fig:c3_550_horizontal_emittance}, while limiting the increase in the BIB and the broadening of the luminosity spectrum~\cite{supplemental}.

Based on the results presented above, we propose a new set of beam parameters for \CCCnospace-550, which we refer to as Parameter Set 2 (PS2),  by modifying the horizontal and vertical emittances to $\epsilon_{x}^{*}=1000 \ \mathrm{nm}$ and $\epsilon_{y}^{*}=12 \ \mathrm{nm}$, respectively, and introducing a vertical waist shift of  $w_{y}=0.8\sigma_{z}^{*}=80$ \textmu m.

\subsection{Parameter Optimization for \CCCnospace-250}
\label{subsec:opt_c3-250}

\begin{figure*}[h!]
\subfloat[\centering\label{fig:c3-250-waist-shift}]{%
    \hspace*{-0.9cm}
    \includegraphics[scale=0.41]{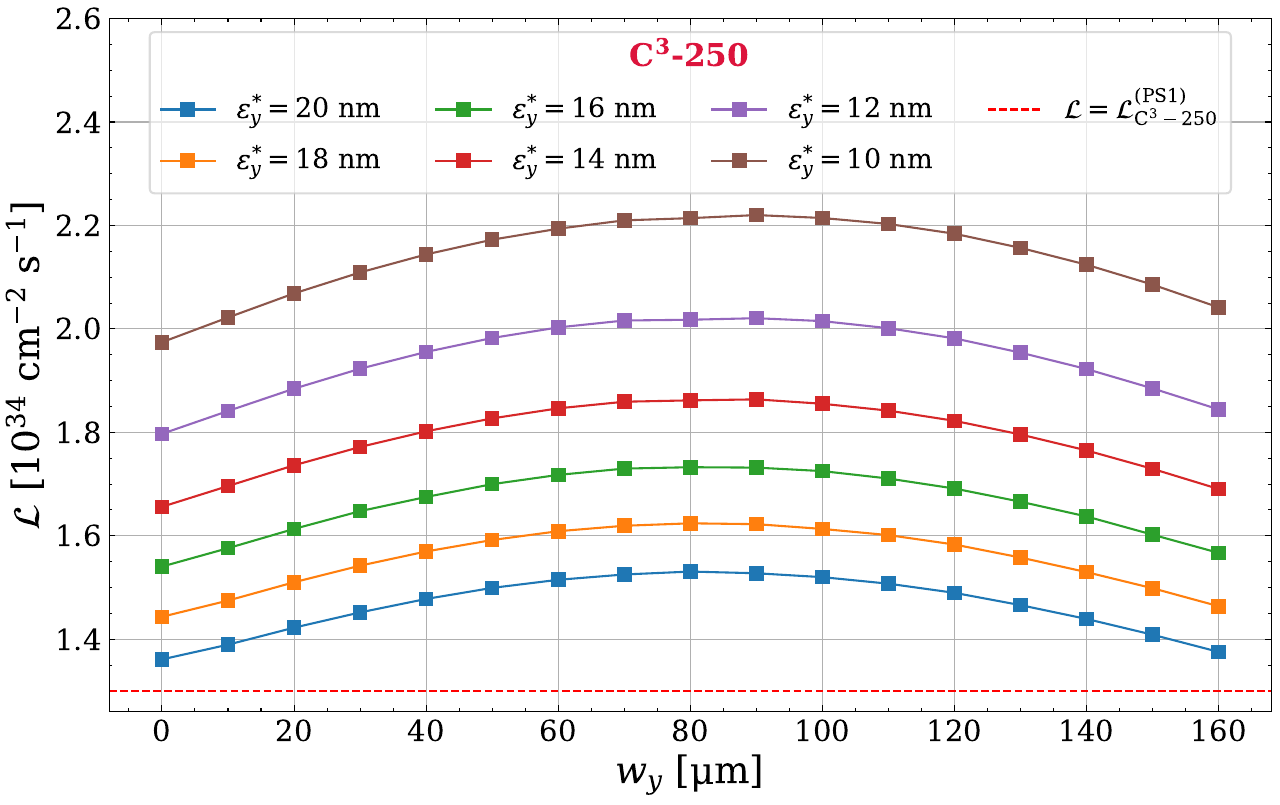}
}
\subfloat[\centering\label{fig:c3-250-vertical-emittance}]{%
    \includegraphics[scale=0.41]{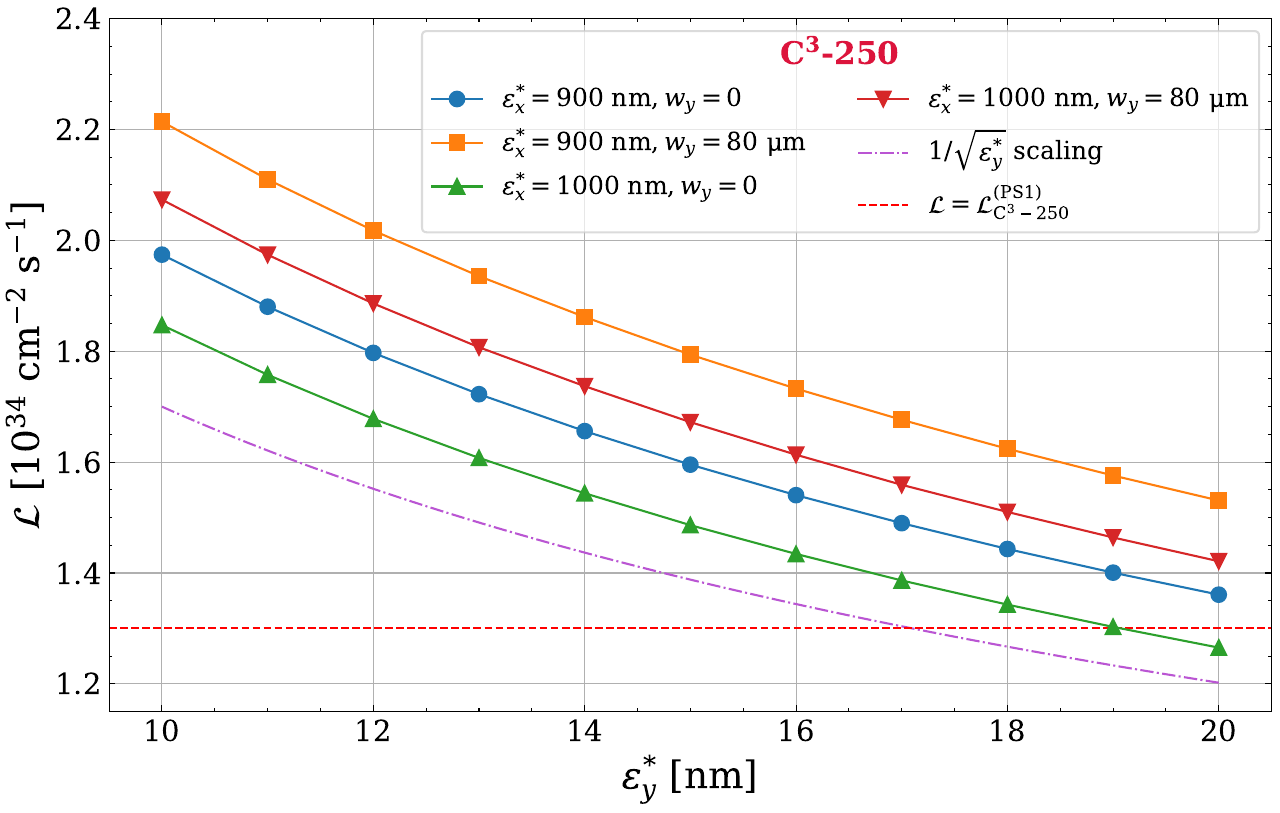}
}
\caption{Luminosity for \CCCnospace-250 as a function of (a) the vertical waist shift $w_y$ and (b) the vertical emittance $\epsilon_y^{*}$. All other beam parameters are kept to their nominal values. In (b), the scaling of the luminosity in the absence of beam-beam interactions $1/\sqrt{\epsilon_{y}^{*}}$ is also given. In both plots, the horizontal red dashed line indicates the target luminosity for \CCCnospace-250.}
\label{fig:lumi_scans_C3_250}
\end{figure*}

Similar to \CCCfivenospace, the luminosity for \CCCtwo can be increased by introducing a vertical waist shift $w_y$. In Figure~\ref{fig:c3-250-waist-shift}, the luminosity for \CCCtwo is shown as a function of $w_y$ for various vertical emittances $\epsilon_{y}^{*}$. Across all scanned $\epsilon_{y}^{*}$ values, the maximum luminosity is achieved for vertical waist shifts of around 80 \textmu m, indicating that the chosen value of $w_y$ for \CCCfive is also optimal for the 250 GeV case. Additionally, the target luminosity for \CCCtwo is exceeded in all cases.

In order to evaluate the effect of the increase in the horizontal emittance proposed in PS2 for \CCCfivenospace, we scan the luminosity for \CCCtwo as a function of vertical emittance for horizontal emittances of $900$ and $1000 \ \mathrm{nm}$ and for waist shifts of $0$ and 80 \textmu m. The results are shown in Figure~\ref{fig:c3-250-vertical-emittance}. We observe that the PS2 configuration of $\epsilon_{x}^{*} = 1000 \ \mathrm{nm}, w_y =  80$ \textmu m achieves the highest luminosities second to the $\epsilon_{x}^{*} = 900 \ \mathrm{nm}, w_y =  80$ \textmu m case, which, however, suffers from larger BIB rates. Since the PS2 configuration for \CCCfive still achieves luminosities far exceeding the target for \CCCtwonospace, we conclude that this parameter set can be adopted for both center-of-mass energy stages of \CCCnospace.

 We summarize the PS2  parameters for \CCC in Table~\ref{tab:beam_params_c3_new}  next to the corresponding parameters for the baseline scenario PS1. The table also includes a comparison of the values for various luminosity and BIB-related quantities. Most notably, this parameter set results in an increase of the instantaneous luminosity from $1.35 ~ (1.70)$ to $1.90 ~ (2.40) \cdot 10^{34} \ \mathrm{cm}^{-2} \ \mathrm{s}^{-1}$ for \CCCnospace-250 (550), which amounts to an improvement of around $ 40 \%$ in both cases. At the same time, the $\langle \Upsilon \rangle$ parameter remains roughly the same, indicating that the BIB is maintained at approximately the same levels.

\begin{table*}[!htbp]
    \centering
      \caption{Baseline (PS1) scenario and new proposed set of beam parameters (PS2) for \CCC at 250 and 550 GeV. Any parameters not mentioned here retain the values introduced in Tables~\ref{tab:beam_params},~\ref{tab:lumi_bkg_params}.}
    \label{tab:beam_params_c3_new}
    \hspace*{-0.4cm}
\begin{tabular}{l c |c c|| c  c}
\hline
Parameter & Symbol [unit] & $\mathrm{C}^3$-250 (PS1) & $\mathrm{C}^3$-250 (PS2) & $\mathrm{C}^3$-550 (PS1) & $\mathrm{C}^3$-550 (PS2) \\ \hline 
\hline Center-of-mass Energy & $\sqrt{s_0}$ [GeV]  & \multicolumn{2}{c||}{250} & \multicolumn{2}{c}{550} \\
RMS bunch length & $\sigma_z^{*}$ [\textmu m]  & \multicolumn{2}{c||}{100} & \multicolumn{2}{c}{100} \\
Horizontal beta function at IP & $\beta_x^{*}~[\mathrm{mm}]$ & \multicolumn{2}{c||}{12} & \multicolumn{2}{c}{12} \\
Vertical beta function at IP & $\beta_y^{*}~[\mathrm{mm}]$ & \multicolumn{2}{c||}{0.12} & \multicolumn{2}{c}{0.12} \\
Normalized horizontal emittance at IP & $\epsilon_x^{*}~[\mathrm{nm}]$ & 900 & 1000 & 900 & 1000 \\
Normalized vertical emittance at IP & $\epsilon_y^{*}~[\mathrm{nm}]$ & 20 & 12 &   20 & 12 \\
RMS horizontal beam size at IP & $\sigma_x^{*}~[\mathrm{nm}]$   & 210 & 221 & 142 & 149 \\
RMS vertical beam size at IP & $\sigma_y^{*}~[\mathrm{nm}]$ & 3.1 & 2.4 & 2.1 & 1.6\\
Vertical waist shift & $w_y$ [\textmu m] & 0 & 80 & 0 & 80 \\ \hline
Geometric Luminosity  & $\mathcal{L}_{\mathrm{geom}}$ $\left[10^{34}  \mathrm{~cm}^{-2} \mathrm{~s}^{-1}\right]$ & 0.75 & 0.92 &  0.93 & 1.14 \\
Horizontal Disruption & $D_x$ & 0.32 & 0.29 & 0.32 & 0.29 \\
Vertical Disruption &  $D_y$ &  21.5 &  26.5 &  21.5 & 26.5 \\
Average Beamstrahlung Parameter & $\langle \Upsilon \rangle$   & 0.065 & 0.062 & 0.21 & 0.20 \\ 
Total Luminosity & $\mathcal{L}$ $\left[10^{34}  \mathrm{~cm}^{-2} \mathrm{~s}^{-1}\right]$ & 1.35 & 1.90 & 1.70 & 2.40 \\
Peak luminosity fraction & $\mathcal{L}_{0.01}/\mathcal{L}$ $[\%]$ & 73 & 74 &  52 & 54 \\ 
Enhancement Factor&  $H_D$  & 1.8 & 2.1 & 1.8 & 2.1 \\
Average Energy loss & $ \delta_{E}$ $[\%]$ & 3.3 & 3.1 & 9.6 & 9.0 \\
Photons per beam particle & $n_{\gamma}$ & 1.4 & 1.3 & 1.9  & 1.8\\
Average Photon Energy fraction & $\langle E_{\gamma}/E_{0} \rangle$ $[\%]$  & 2.5 & 2.4 &  5.1 & 5.0 \\
Number of incoherent particles/BX & $N_{\mathrm{incoh}}$ [$10^{4}$]  & 4.7 & 5.9 & 12.6 & 15.5 \\
Total energy of incoh. particles/BX & $E_{\mathrm{incoh}}$ [TeV] & 58 & 71 & 644 & 768 \\
\hline \hline
\end{tabular}
\end{table*}

\subsection{Luminosity dependence on beta function}

\begin{figure*}[h!]
\subfloat[\centering \label{fig:lumi_scan_C3_betax}]{%
\hspace*{-0.9cm}
    \includegraphics[scale=0.39]{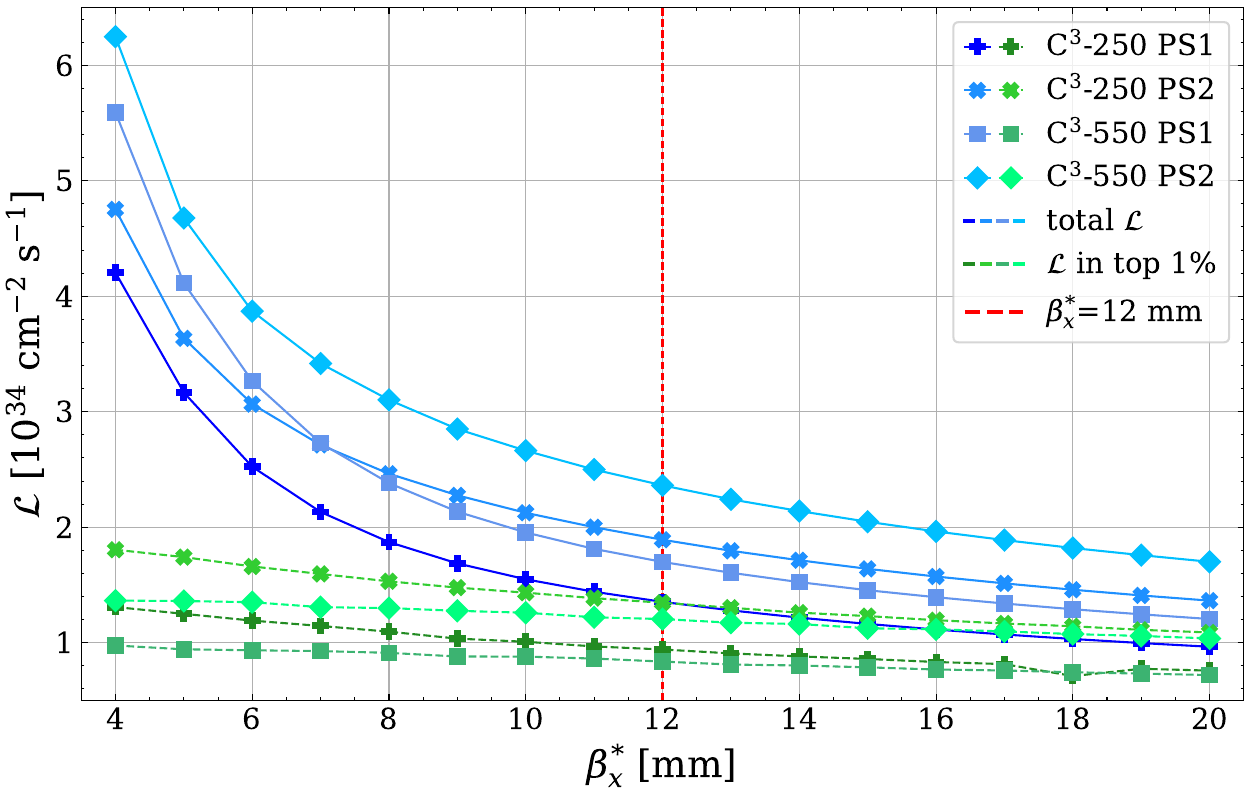}
}
\subfloat[\centering \label{fig:lumi_scan_C3_betay}]{%
    \includegraphics[scale=0.39]{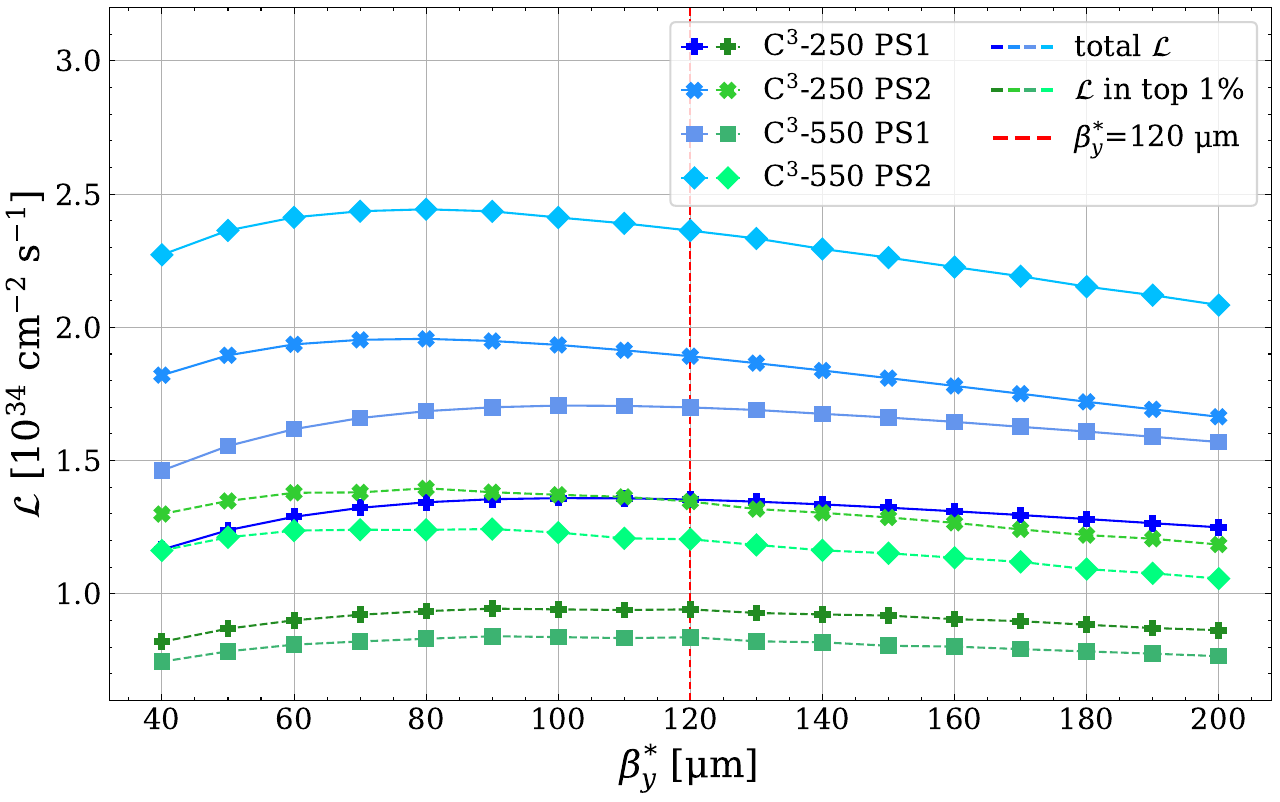}
}
\caption{Luminosity scans for the two parameter sets PS1 and PS2 for \CCCnospace-250 and \CCCnospace-550  as a function of (a) the horizontal $\beta_{x}^{*}$ and (b) the vertical $\beta_{y}^{*}$ beta function at the IP. The lines colored in hues of blue correspond to the total instantaneous luminosity, integrated over all $\sqrt{s}$ values of the colliding particles, whereas the ones in hues of green correspond to the instantaneous luminosity in the top $1 \%$ of $\sqrt{s}$. The red dashed lines correspond to the values of the beta functions chosen for both PS1 and PS2. }
\label{fig:lumi_scan_C3_beta}
\end{figure*}

In the previous analysis, we examined the dependence of the luminosity for \CCC on the horizontal and vertical emittance $\epsilon_{x}^{*}, \epsilon_{y}^{*}$, the bunch length $\sigma_{z}^{*}$ and the vertical waist shift $w_y$. As can be seen from Eq.~\eqref{eq:sigmaxy}, the luminosity also depends on the horizontal and vertical beta functions at the IP  $\beta_{x}^{*},\beta_{y}^{*}$, which affect the RMS bunch sizes in the transverse plane. The lower limits of the beta functions depend on how tightly the bunches can be focused by the Final Focus (FF) system. In order to avoid further assumptions on the tolerances for the \CCC FF system, these parameters were not included in the optimization process, however we examine their effect on the luminosity in Figures~\ref{fig:lumi_scan_C3_betax} and \ref{fig:lumi_scan_C3_betay}, which show the luminosity for \CCC at 250 and 550 GeV, for both parameter sets PS1 and PS2, as a function of $\beta_{x}^{*}$ and $\beta_{y}^{*}$ respectively.

\begin{figure*}[h!]
\subfloat[\centering\label{fig:beta_y_shape_around_IP}]{%
\hspace*{-0.9cm}
    \includegraphics[scale=0.39]{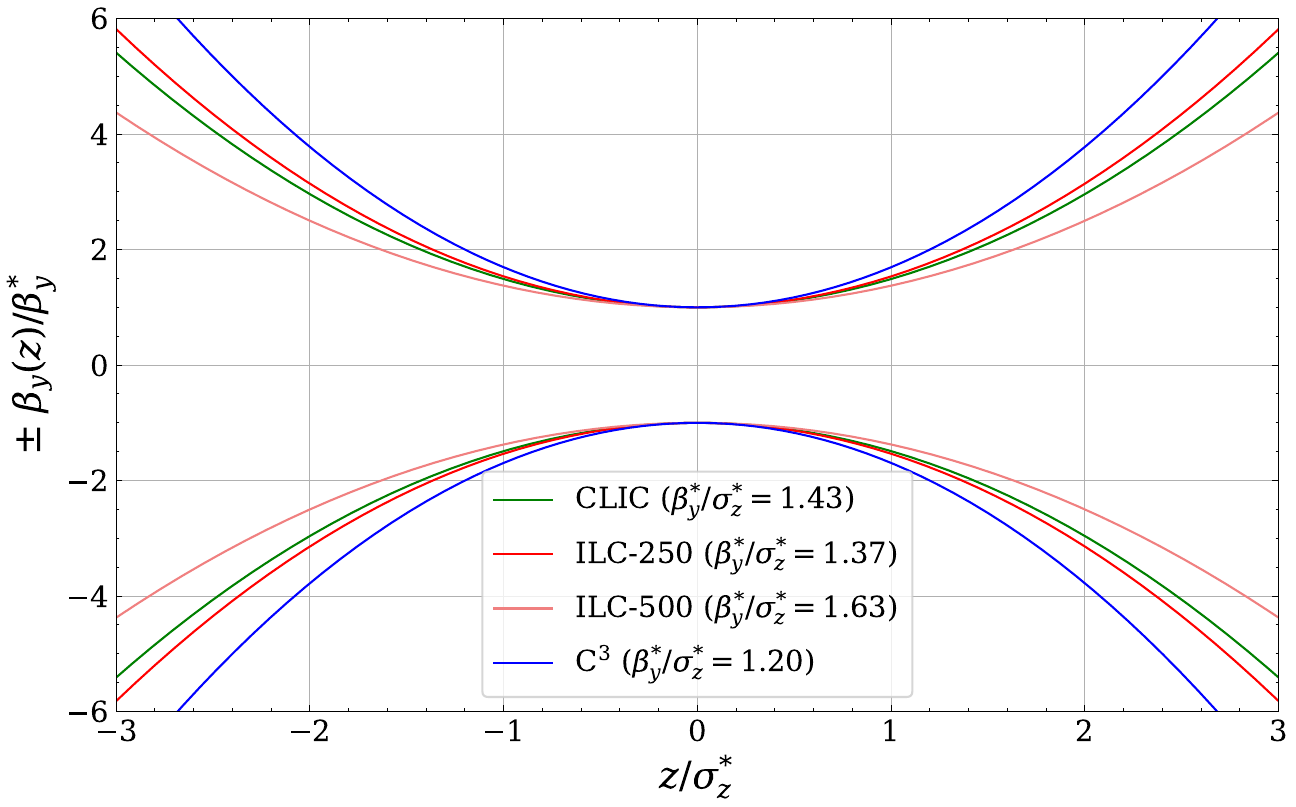}
}
\subfloat[\centering\label{fig:offset_y_scan_norm_lumi}]{%
    \includegraphics[scale=0.39]{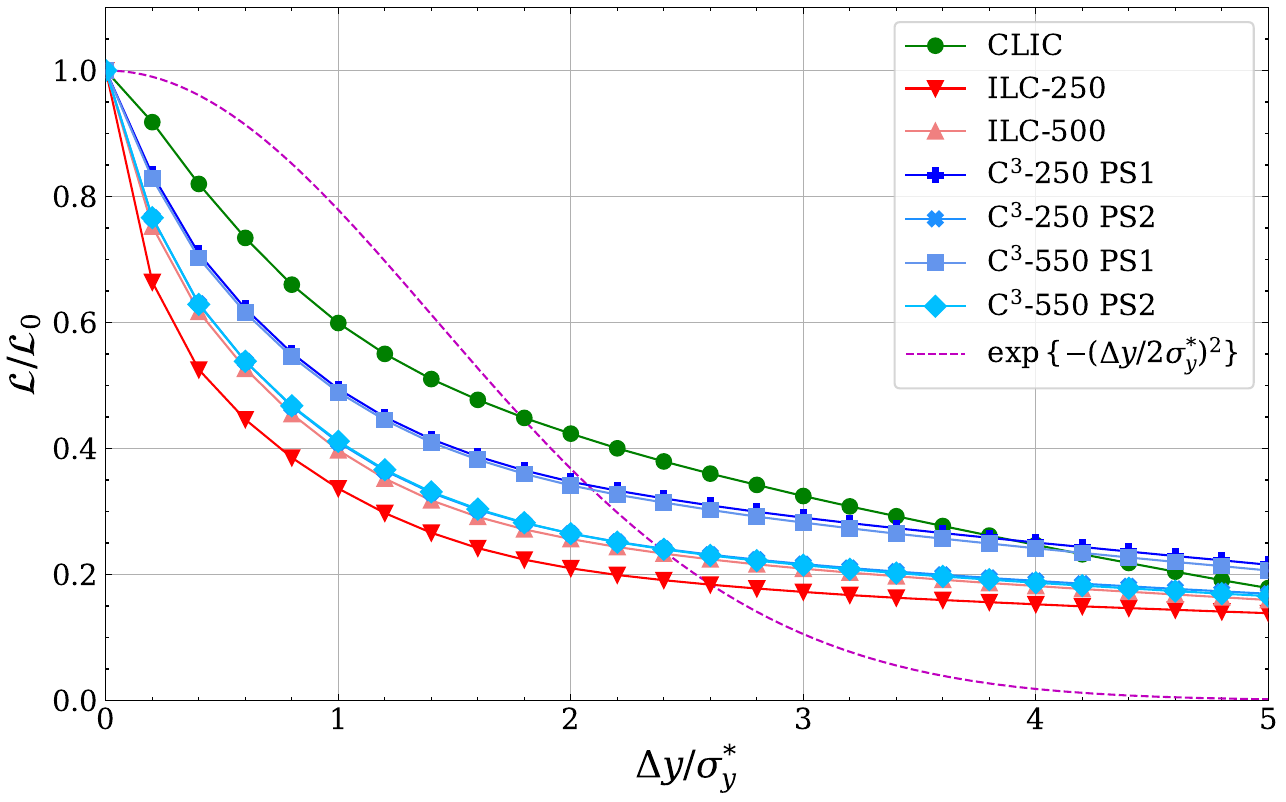}
}
\caption{(a) Normalized beta function $\beta_{y}/\beta_{y}^{*}$ as a function of the relative longitudinal distance $z/\sigma_{z}^{*}$ around the IP and (b) luminosity normalized with respect to its value when assuming zero offset $\mathcal{L}/\mathcal{L}_0$ as a function of the relative vertical beam offset $\Delta y/\sigma_{y}^{*}$.}
\label{fig:hourglass_effect_and_offset_y_scan}
\end{figure*}

From Figure~\ref{fig:lumi_scan_C3_betax}, we observe that the total luminosity increases significantly at smaller values of the horizontal beta function, with the values for PS2 being consistently higher than the respective ones for PS1. The picture changes when looking at the luminosity in the top $1 \%$, which has a much more modest increase at lower $\beta_{x}^{*}$. This stems from the increase in beamstrahlung at lower $\beta_{x}^{*}$, as can be seen from Eq.~\eqref{eq:upsilon}, leading to a broader luminosity spectrum and, thus, a smaller percentage of the luminosity remaining in the top $1 \%$ of center-of-mass energies. A different effect is observed in Figure~\ref{fig:lumi_scan_C3_betay} under variations in the vertical beta function, with the luminosity increasing at smaller values of $\beta_{y}^{*}$ and decreasing for even smaller ones. This is a result of the hourglass effect introduced in Section~\ref{sec:physical_quantities}, according to which, as $\beta_{y}^{*}$ --- and, consequently, the beam spot size at the IP --- decreases, the rate at which the beam broadens around the IP increases. This is demonstrated in Figure~\ref{fig:beta_y_shape_around_IP}. Since the vertical beta function is of the same order of magnitude as the bunch length, this broadening affects the bunch size around the IP, leading to luminosity degradation at small values of  $\beta_{y}^{*}$.

Overall, we conclude that luminosity gains through beta function decreases are conceivable both in the horizontal and vertical direction. In the latter case, the gains are attenuated by the hourglass effect to the few percent level, while in the former significant total luminosity enhancement comes at the cost of increases in the BIB and degradation of the luminosity spectrum. Understanding up to what extent the background levels can increase without compromising detector performance requirements, as well as how exactly the luminosity spectrum degradation influences the expected precision in the measurement of physical observables of interest, would require further investigation. For these reasons, and pending more advanced studies of the FF system for \CCC in order to evaluate the implications of further beam focusing on the accelerator side, we have decided to retain the horizontal and vertical beta functions at their current values of 12 mm and 120 \textmu m respectively.

\vspace*{-0.2cm}

\subsection{Luminosity dependence on beam offset}
\label{subsec:offset}

In the above analysis, the idealized assumption of head-on collisions of the bunches was made in both the horizontal and vertical directions. In the case where the two beams are not perfectly aligned, the bunches can intersect at the IP with non-zero horizontal ($\Delta x$) and/or vertical ($\Delta y$) offsets, leading to luminosity degradation. For rigid, non-interacting bunches with Gaussian charge densities, the effect of beam-beam offsets on the luminosity would simply be

\begin{equation}
     \mathcal{L}_{\mathrm{inst}}=H_{D}\frac{N_{e}^2 n_{b} f_{r}}{4\pi \sigma_{x}^{*}\sigma_{y}^{*}} \cdot \exp{\left \{-\left(\dfrac{\Delta x}{2\sigma_{x}^{*}} \right)^2 - \left(\dfrac{\Delta y}{2\sigma_{y}^{*}} \right)^2 \right \} }
     \label{eq:lumi_offset_analytical}
\end{equation}

\noindent leading to a rapid loss of luminosity with increasing offset. When beam-beam interactions are taken into account, the offset dependence is modified as shown in Figure~\ref{fig:offset_y_scan_norm_lumi} for $\Delta y$, where the luminosity for CLIC, ILC, and \CCC (for both parameter sets) has been estimated for different offsets in the horizontal and vertical direction.

In general, the luminosity degrades faster than the rigid-beams case at small offsets, due to the large disruptions causing the collisions to become unstable. This effect is known as the kink instability~\cite{Chin:1987ct, Chen:1988ng}. At larger offset values, the attractive interactions between the two oppositely charged bunches allow for higher luminosity retention compared to the rigid-beams case.  The dependence of the luminosity degradation under increases in the vertical offset on the disruption parameter is verified by juxtaposing the curves in Figure~\ref{fig:offset_y_scan_norm_lumi} with the vertical disruption parameter $D_y$ values in Tables~\ref{tab:lumi_bkg_params}, \ref{tab:beam_params_c3_new}. CLIC has the smallest value for $D_y$, followed by \CCC for the PS1 beam parameters\footnote{Note that the disruption parameter is independent of the center-of-mass energy under constant $\epsilon^{*},\beta^{*}$ and so has the same value for \CCC at both 250 and 550 GeV.}. For the PS2 beam configuration, $D_y$ is larger than for PS1 and has almost the same value as for ILC-500. Finally, ILC at 250 GeV has the largest $D_y$ value. The same trend is reflected in the luminosity curves of Figure~\ref{fig:offset_y_scan_norm_lumi}.

Finally, Figure~\ref{fig:offset_y_scan_norm_lumi} indicates that a beam alignment at the IP of sub-nm level precision is required in order to maintain high instantaneous luminosities, since a vertical offset of one $\sigma_{y}^{*}$ can lead to a $40 - 65\%$ luminosity degradation, depending on the collider and the beam parameter configuration. This emphasizes the importance of highly precise alignment of the accelerator components.

Offsets in vertical beam position can result from many sources including rf amplitude and phase jitter, beam time of arrival, beam current stability, and alignment jitter. Sensitivity to these sources requires detailed simulations of the main linac and beam delivery system. Similar work performed for CLIC~\cite{Schulte:2010bga,Gohil:2020tzn} can provide initial guidance on some of these sensitivities. For example, the target rf amplitude and phase are achievable~\cite{Nanni:2023yne,Liu:2024vej} and within the requirements anticipated for CLIC.

For \CCC specifically, alignment of the accelerating structures and quadrupole magnets in the main linac and the BDS at the 10 \textmu m level is necessary to maintain beam steering~\cite{Nanni:2023yne}. This is made challenging due to vibrations from human activities and seismic disturbances and, most importantly, from the nucleate boiling and the subsequent vapor flow of the liquid nitrogen used to cool the copper cavities. Various techniques could be utilized to address this, including the Rasnik  3-point optical alignment system~\cite{vanderGraaf:2021wxg,vanderGraaf:2023oar}. At the IP, for improved performance and correction of alignment jitter from effects such as ground motion, the sub-nm level alignment can be achieved with a beam-based active feedback system~\cite{Burrows:2016ghg}, such as FONT3~\cite{Burrows:2005jf} which can correct for beam-beam offsets within a few \CCC bunch crossings.

\subsection{Power consumption considerations}
\label{subsec:power_consumption}

Further optimizations can be explored by allowing the remaining beam parameters, namely the bunch charge $Q$, the number of bunches per train $n_b$, and the train repetition rate $f_r$, to vary from their baseline values. In our previous analysis, the decision was made to keep the bunch charge at its nominal value of 1 nC, in order to stay within the tolerances set by the current \CCC rf design, most notably the aperture size and the breakdown rate requirements~\cite{Vernieri_2023}. At the same time, $n_b$ and $f_r$ were kept constant in order to maintain a low beam power, cf. Equation~\eqref{eq:beam_power}, and, by extension, not increase the power requirements on the rf and cryogenics systems of the main linac. This is crucial for maintaining the overall site power within reasonable limits and achieving  \CCCnospace 's envisioned sustainable operation design, as laid out in~\cite{c3_sust}.

In the same reference, the environmental impact of \CCC is evaluated in the context of energy requirements and carbon footprint for construction and operation, and power-saving scenarios are proposed, without sacrifices in the instantaneous luminosity. These scenarios entail a doubling of $n_b$ by either doubling the flat top, i.e. the duration of maximum constant acceleration gradient, or by halving the bunch spacing. In both cases, the doubling of $n_b$ is compensated by a decrease in $f_r$ from 120 to 60 Hz, resulting, overall,   in the same luminosity. One could combine both scenarios to achieve quadruple number of bunches per train compared to the baseline scenario. This translates to a 100 (300) $\%$ increase in the instantaneous luminosity for $f_{r} = 60 \ (120) \ \mathrm{Hz}$. We summarize these scenarios in Tables~\ref{tab:power_opt_scenarios_c3_250} and \ref{tab:power_opt_scenarios_c3_550} for \CCCtwo and \CCCfive respectively.

\begin{table*}[!htbp]
    \centering
    \caption{Beam configuration scenarios for \CCCnospace-250 which include modifications in the bunch spacing $\Delta t_{b}$, the number of bunches per train $n_{b}$ and/or the train repetition rate $f_{r}$. The last three columns give the instantaneous luminosity for the PS1 and PS2 parameter sets, as well as the estimated total site power, in each case.}
    \label{tab:power_opt_scenarios_c3_250}
    \begin{tabular}{c c|c|c|c|c|c|c}
    \multicolumn{5}{c|}{} & \multicolumn{2}{c|}{$\mathcal{L}$ $\left[ 10^{34} \ \mathrm{cm}^{-2} \ \mathrm{s}^{-1}\right]$} & \multicolumn{1}{c}{$P_{\mathrm{site}}$ [MW]} \\
    \hline \hline 
        Scenario & Flat top [ns] & $\Delta t_{b}$ [ns] & $n_b$ & $f_r$ (Hz) & \CCCnospace-250 (PS1) & \CCCnospace-250 (PS2) & Both scenarios \\ \hline 
        Baseline & 700 & 5.26 & 133 & 120 &  1.35 & 1.90 & 150  \\ 
        Double flat top & 1400 & 5.26 & 266 & 60  & 1.35 & 1.90  & 125   \\ 
        Halve bunch spacing & 700 & 2.63 & 266 & 60  & 1.35 & 1.90 & 129   \\ 
        Combined - half rep. rate & 1400 & 2.63 & 532 & 60  &2.70 & 3.80 & 154   \\ 
        Combined - nominal rep.rate & 1400 & 2.63 & 532 & 120  &5.40 & 7.60 & 180  \\ \hline \hline 
    \end{tabular}
\end{table*}

\begin{table*}[!htbp]
    \centering
    \caption{Beam configuration scenarios for \CCCnospace-550 which include modifications in the bunch spacing $\Delta t_{b}$, the number of bunches per train $n_{b}$ and/or the train repetition rate $f_{r}$. The last three columns give the instantaneous luminosity for the PS1 and PS2 parameter sets, as well as the estimated total site power, in each case.}
    \label{tab:power_opt_scenarios_c3_550}
    \begin{tabular}{c c|c|c|c|c|c|c}
    \multicolumn{5}{c|}{} & \multicolumn{2}{c|}{$\mathcal{L}$ $\left[ 10^{34} \ \mathrm{cm}^{-2} \ \mathrm{s}^{-1}\right]$} & \multicolumn{1}{c}{$P_{\mathrm{site}}$ [MW]} \\
    \hline \hline 
        Scenario & Flat top [ns] & $\Delta t_{b}$ [ns] & $n_b$ & $f_r$ (Hz) & \CCCnospace-550 (PS1) & \CCCnospace-550 (PS2) & Both scenarios \\ \hline 
        Baseline & 250 & 3.50 & 75 & 120 &  1.70 & 2.40 & 175   \\ 
        Double flat top & 500 & 3.50 & 150 & 60  & 1.70 & 2.40 & 144  \\ 
        Halve bunch spacing & 250 & 1.75 & 150 & 60  & 1.70 & 2.40 & 149   \\ 
        Combined - half rep. rate & 500 & 1.75 & 300 & 60 &  3.40 & 4.80 & 180   \\ 
        Combined - nominal rep.rate & 500 & 1.75 & 300 & 120  &6.80 & 9.60& 212   \\ \hline \hline 
    \end{tabular}
\end{table*}

For the baseline scenario in Tables~\ref{tab:power_opt_scenarios_c3_250} and \ref{tab:power_opt_scenarios_c3_550}, a distribution of the power for the main linac among the rf system and the cryoplant of 40 (65) MW and 60 MW, respectively, has been calculated for \CCCtwo(550) assuming current industrial technologies, with an additional 50 MW for the accelerator complex beyond the main linac. This amounts to a total site power of 150 MW at 250 GeV and 175 MW at 550 GeV. Reducing the repetition rate while extending the flat top or shortening the bunch spacing would result in decreased thermal dissipation in the main linac, thus reducing the overall power consumption. The estimates of the total site power for the various scenarios have been extracted following the methodology in~\cite{c3_sust}. Additional power savings stemming from improvements in the rf source efficiency and the utilization of pulse compression have not been assumed here.


We note that the scenarios above indicate that significant luminosity gains are achievable through modifications in $n_b$ and $f_r$, with only moderate increases in the site power consumption.
Nevertheless, detailed studies are warranted in order to guarantee the feasibility of these scenarios, both in terms of accelerator design, including high-gradient testing in order to determine whether doubling the flat top is achievable, as well as detector performance, most notably evaluating detector occupancy when increasing the train duration or reducing the bunch spacing, which lead to higher fluxes of background particle hits.

\section{Comparison of various linear collider proposals}
\label{sec:comparison}

The luminosity- and BIB-related quantities for CLIC, ILC, and \CCC  are summarized in Tables~\ref{tab:lumi_bkg_params} and \ref{tab:beam_params_c3_new}. All these colliders use flat beams of similar dimensions and bunch charges and achieve luminosities of $1.3-1.8 \cdot 10^{34} \ \mathrm{cm}^{-2} \  \mathrm{s}^{-1}$, with the updated \CCC configuration reaching even higher values. The average energy loss due to beamstrahlung is at the $3-10\%$ level, with the lowest (highest) value achieved for ILC-250 (\CCCnospace-550). The average Beamstrahlung parameter is $\langle \Upsilon \rangle \lesssim 0.2$, meaning that the dominant background process is incoherent pair production. The number of such incoherent pair particles produced is of the order of $10^{4}-10^{5}$, with larger numbers for the higher center-of-mass energy runs of ILC and \CCCnospace.

\begin{figure*}[htbp]
    \centering
    \includegraphics[width=0.86\columnwidth]{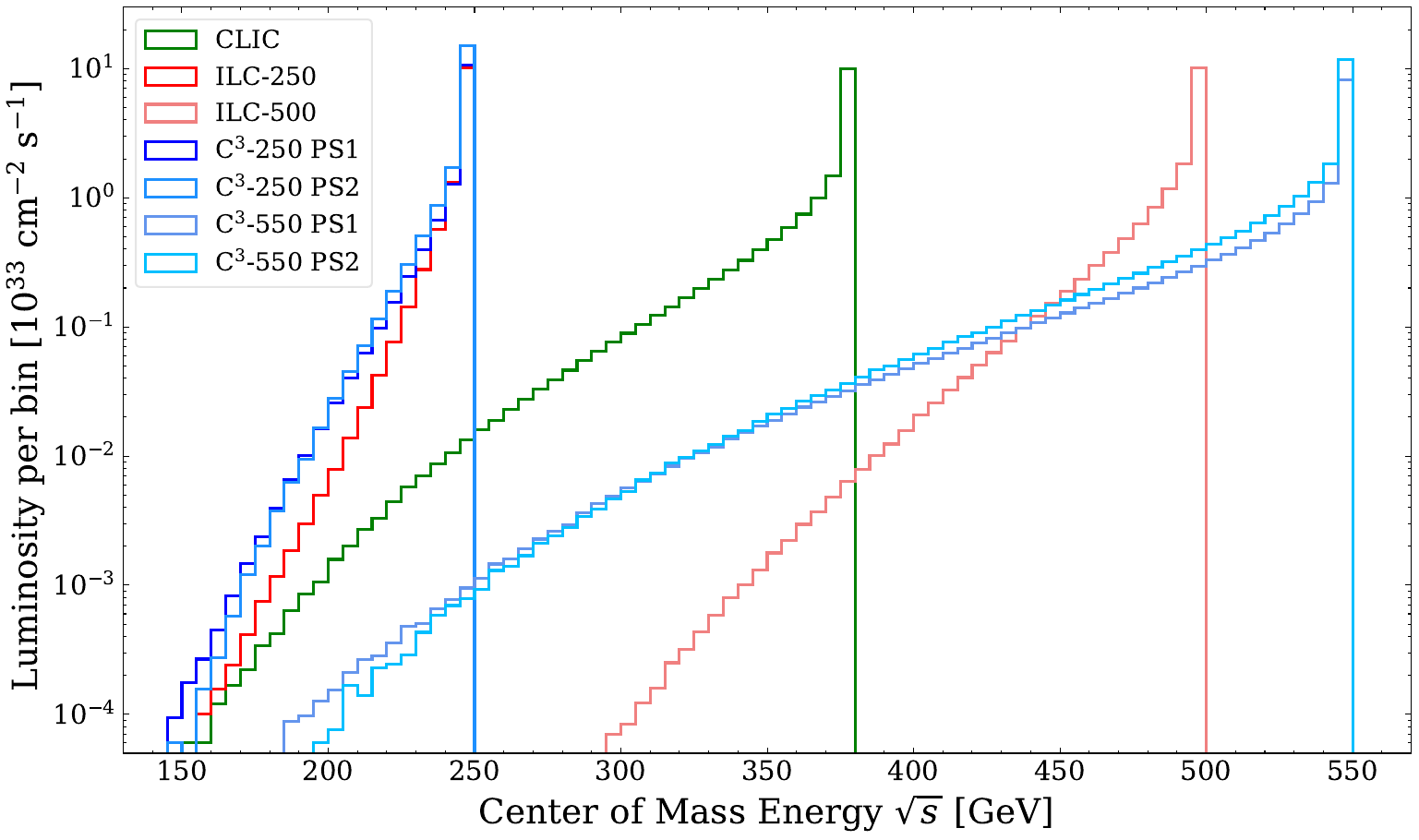}
    \caption{Luminosity spectra for the linear collider proposals under consideration here, as obtained from GUINEA-PIG simulations using the beam parameters from Tables~\ref{tab:beam_params}, \ref{tab:beam_params_c3_new}.}
    \label{fig:lumi_spectra_130_to_570GeV}
\end{figure*}

The proposed colliders in Table~\ref{tab:beam_params} can also be compared in terms of their luminosity spectra, which indicate how broad the center-of-mass energy distributions of the colliding particles are, and therefore affect the level of precision to which the four-momenta of initial state particles can be known. Figure~\ref{fig:lumi_spectra_130_to_570GeV} shows the luminosity spectra for the various linear colliders under consideration, obtained from GUINEA-PIG simulations with the beam parameters of Table~\ref{tab:beam_params}. For \CCCnospace, the luminosity spectra for both PS1 and PS2 are shown. All luminosity spectra contain the effects of beamstrahlung and initial energy spread at the IP (before beamstrahlung), but not initial-state radiation (ISR). In all cases, most of the luminosity is contained near the nominal center-of-mass energy $\sqrt{s_0}$, with tails corresponding to contributions from beam particles that lost a significant amount of their initial four-momentum due to Beamstrahlung. For \CCC specifically, one observes that the PS2 beam configuration achieves noticeably higher luminosities at the peak, compared to PS1, whereas the tails are comparable, reaffirming our conclusion that the newly proposed parameter set leads to overall higher luminosities without correspondingly increasing the BIB.

\begin{figure*}[h!]
\subfloat[\centering\label{fig:lumi_spectra_0p5_to_1}]{%
\hspace*{-0.8cm}
    \includegraphics[scale=0.356]{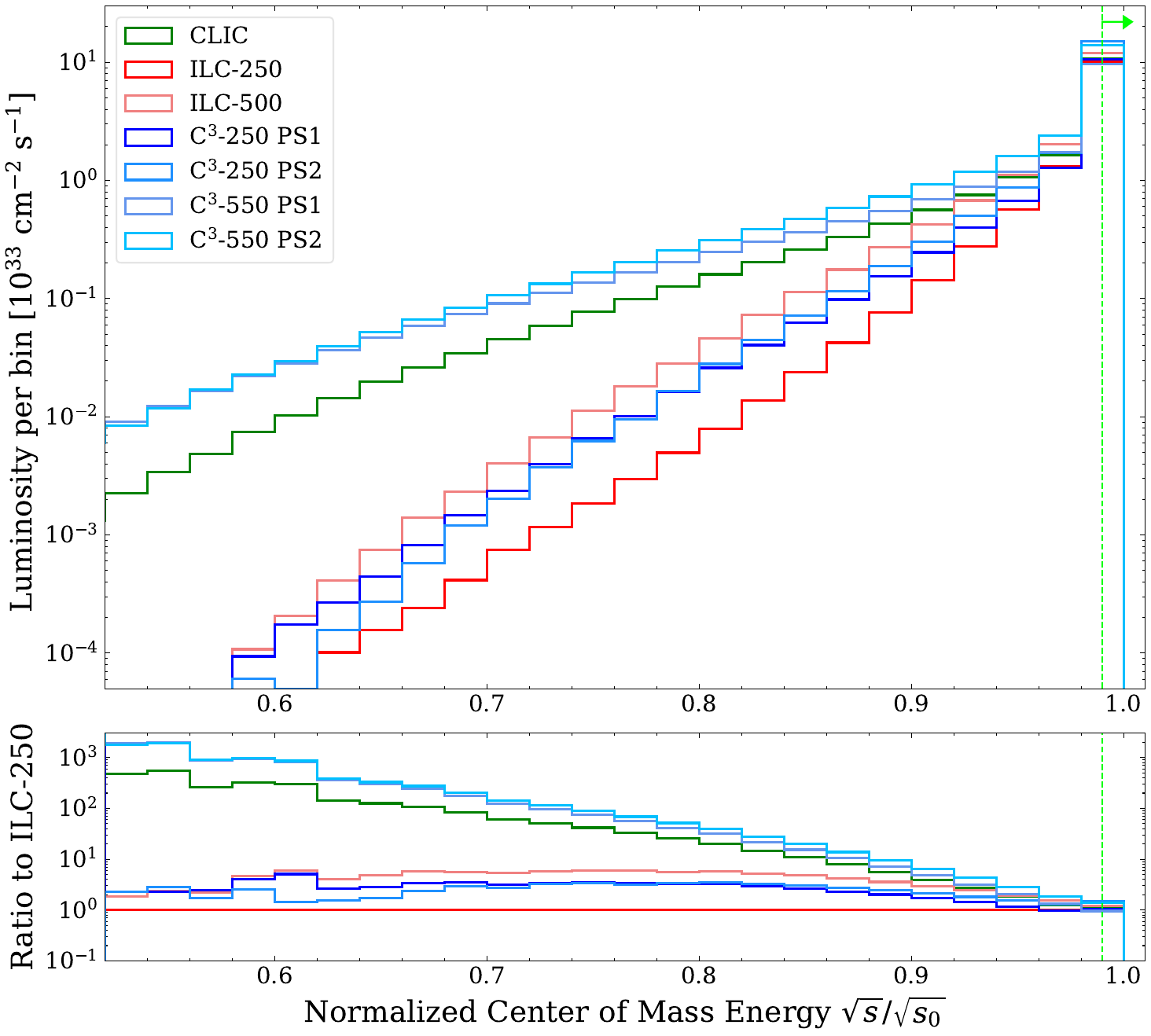}
}
\subfloat[\centering\label{fig:lumi_spectra_0p9_to_1}]{%
    \includegraphics[scale=0.356]{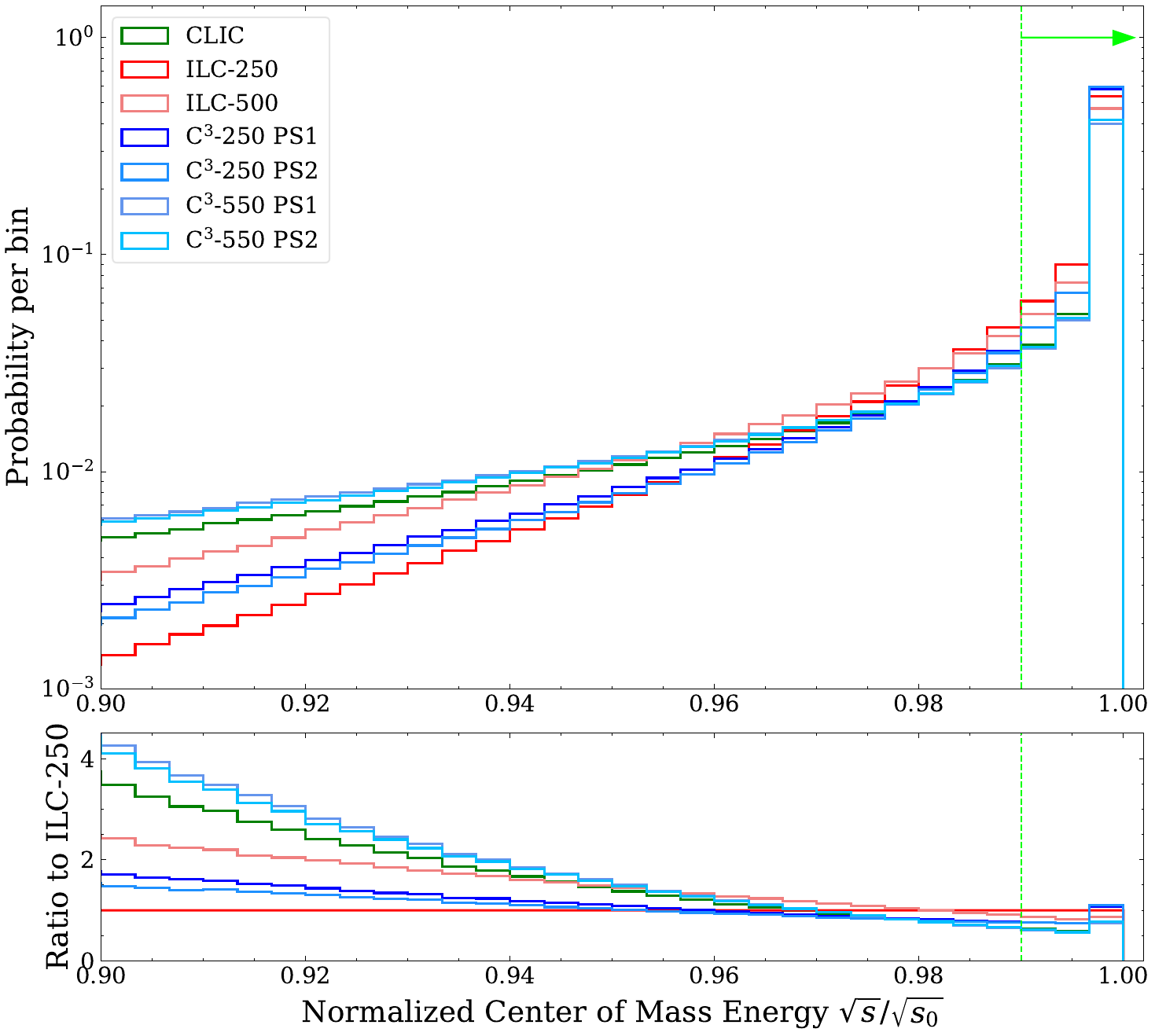}
}
\caption{Luminosity spectra of different linear colliders (top panel) and ratios with respect to the one for ILC-250 (bottom panel) for normalized center-of-mass energies (a)  $\frac{\sqrt{s}}{\sqrt{s_0}}>0.5$ and (b)  $\frac{\sqrt{s}}{\sqrt{s_0}}>0.9$. The vertical green dashed line and horizontal arrow indicate the fraction of the luminosity spectrum that corresponds to center-of-mass energies $\sqrt{s}>0.99 \sqrt{s_0}$.}
\label{fig:lumi_spectra_normalized_com}
\end{figure*}

Further comparison of the luminosity spectra is facilitated by normalizing the center-of-mass energy of each collision $\sqrt{s}$ to its nominal value $\sqrt{s_0}$, as shown in Figure~\ref{fig:lumi_spectra_normalized_com}. In Figure~\ref{fig:lumi_spectra_0p5_to_1}, the luminosity spectra for $\sqrt{s}/\sqrt{s_0} \geq 0.5$ are shown, indicating that \CCCfive has the highest peak luminosity and ILC-250 has the narrowest luminosity spectrum, with the luminosity tails for \CCCfive being up to 3 orders of magnitude larger for $\sqrt{s}/\sqrt{s_0} \simeq 0.5$. In Figure~\ref{fig:lumi_spectra_0p9_to_1} we zoom in closer on the peak of the luminosity spectra. The luminosity fraction in the top $1 \%$ of the center-of-mass energy $\mathcal{L}_{0.01}/\mathcal{L}$ is indicated by the green arrow and is also shown in Tables~\ref{tab:lumi_bkg_params}, \ref{tab:beam_params_c3_new}, with its value ranging between $52 \% - 74 \%$ among the different colliders. We note that ILC-250, C$^3$-250 achieve the highest peak luminosity fractions, with C$^3$-550 achieving the lowest. Generally, higher values of the peak luminosity fraction are more desirable, as they help better constrain the four-momentum of the colliding \ee \  pair and, thus, of the subsequently produced final-state particles. This is crucial for precision measurements, which have been demonstrated to yield improved sensitivities with techniques that utilize such constraints, such as kinematic fits~\cite{List:88030, BECKMANN2010184, Radkhorrami:2021cuy}. It is worth noting that, although values of $\mathcal{L}_{0.01}/\mathcal{L}$ of around $60\%$ have conventionally been set as goals for \ee \ machines operating in the Higgs factory regime~\cite{schulte_beam_beam, CLICdp:2018cto, CLICdp:2018esa}, detailed studies are necessary, using full detector simulations with luminosity spectra of various widths as inputs, in order to fully understand the exact effect of the value of $\mathcal{L}_{0.01}/\mathcal{L}$ on the level of precision with which physics observables of interest can be measured.

\begin{figure*}[h!]
\subfloat[\centering \label{fig:energy_incoh_pairs}]{%
\hspace*{-1.0cm}
    \includegraphics[scale=0.42]{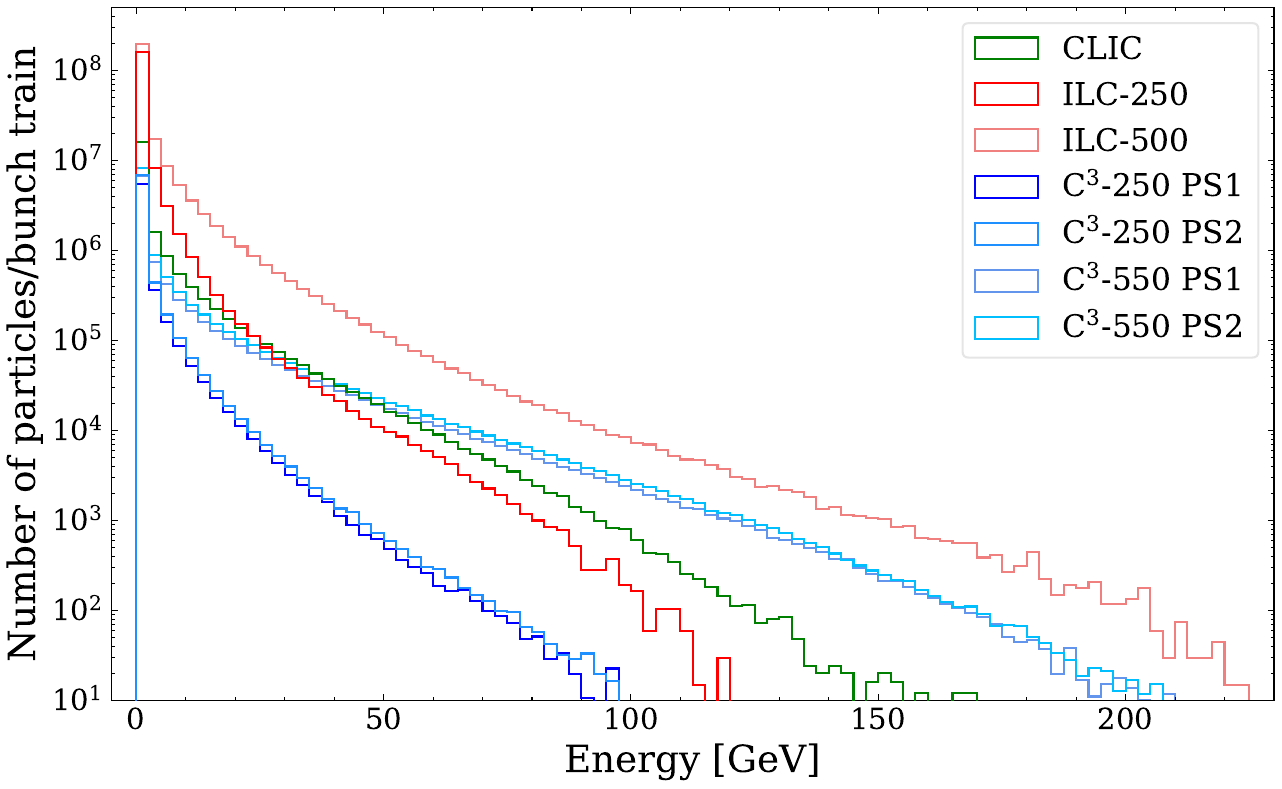}
}
\subfloat[\centering\label{fig:boost_incoh_pairs}]{%
    \includegraphics[scale=0.42]{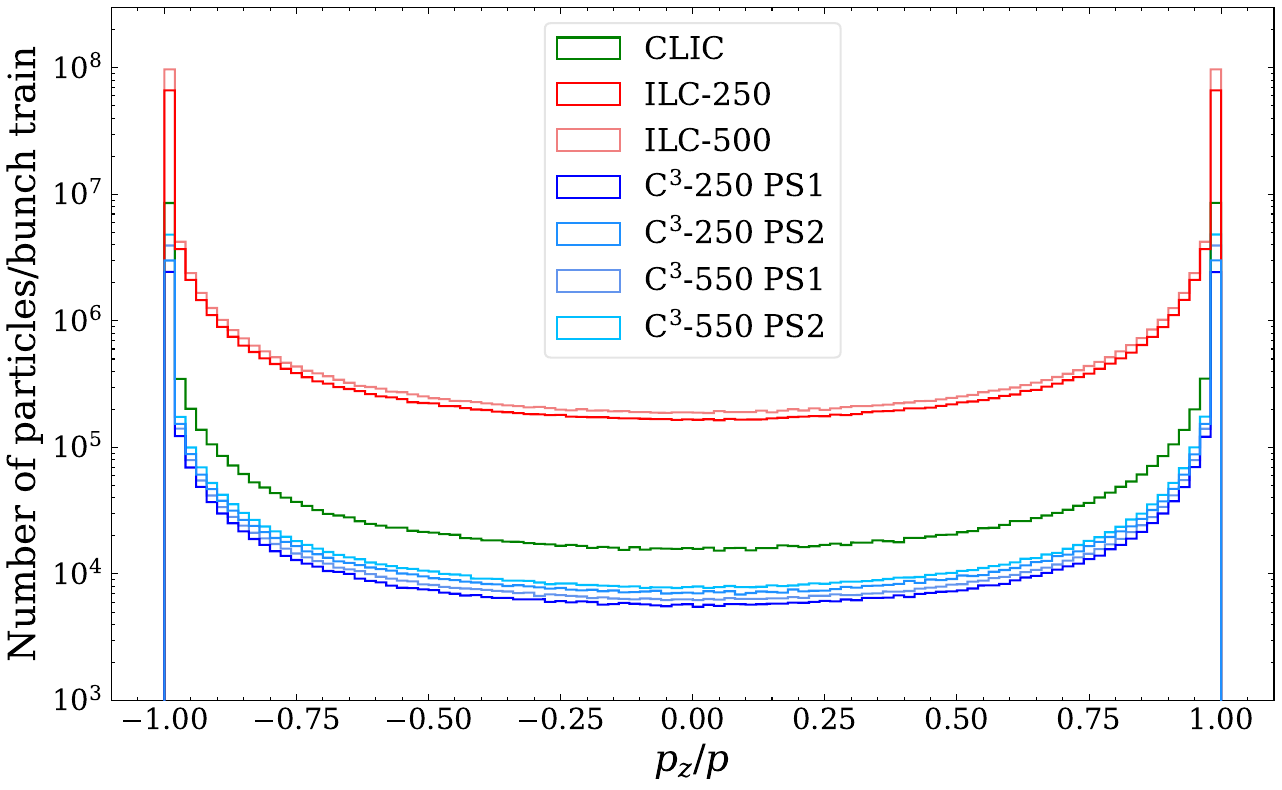}
}
\caption{Distributions of (a) the energy and  (b) the longitudinal boost of the incoherent $e^{+}e^{-}$ pairs for various linear collider proposals. Each distribution has been normalized to the expected number of incoherent pair particles per bunch train $N_{\mathrm{incoh}}\cdot n_{b}$.}
\label{fig:incoh_pairs_spectra}
\end{figure*}

Finally, the distributions of the incoherently produced \ee \ secondaries can be used to characterize, to first order,  the magnitude of the BIB, which is important for detector design purposes. Figure~\ref{fig:incoh_pairs_spectra} shows the distributions of the energy $E$ and the longitudinal boost $p_{z}/p$, where $p$ is the magnitude of the momentum vector and $p_{z}$ is its projection along the beam axis, of the incoherently produced pair particles for the various linear colliders. These distributions have been normalized to the expected number of such pair particles produced over an entire bunch train, assuming a common per-train readout scheme for all colliders. Such a readout scheme has been envisioned for both ILC~\cite{Breidenbach:2021sdo, ILDConceptGroup:2020sfq} and CLIC~\cite{CLICdp:2017vju} detectors and would enable power-pulsing, whereby the recorded hits are buffered on the front-end electronics and read out at the end of a bunch train, with the front-end electronics subsequently powered down in order to reduce power consumption.

Since ILC relies on superconducting rf accelerating technology --- unlike CLIC and \CCCnospace, which utilize normal-conducting cavities --- the train repetition rate $f_r$ is limited to $5-10 \ \mathrm{Hz}$, leading to an order of magnitude larger number of bunches per train. For this reason, the number of background particles per train is the highest for ILC in both center-of-mass energy scenarios, as can be seen from Figure~\ref{fig:incoh_pairs_spectra}. For \CCCnospace, this number is smaller for the 250 GeV compared to the 550 GeV stage, due to the larger $\langle \Upsilon \rangle$ parameter at 550 GeV, with only small differentiation observed between the baseline and new proposed scenarios, indicating that a luminosity enhancement for \CCC is achievable without a significant increase in the beam-beam background rates. In general, the energy and momentum of the incoherent pair particles increase with the center-of-mass energy, whereas the longitudinal boost remains close to unity, in absolute values, indicating that these particles are highly boosted in the forward direction. This fact, in addition to the deflection of the particles in the strong magnetic field of the detectors, leads to most of these background particles traveling outwards from the collision region contained within the beam pipe and thus not reaching sensitive detector components~\cite{schulte_beam_beam, Vogel:2008zza, Dannheim:1443516, Arominski:2704642}. This has important implications for detector design for any \ee \ Higgs factory, since, despite the large number (order of magnitude of $10^6 - 10^8$) of background particles produced in each bunch train, only a small fraction of them --- at the per mille level or less --- contribute to detector occupancy.

\vspace*{-0.3cm}
\section{Conclusions}
\label{sec:conclusions}

We presented an overview of the luminosity and beam-induced background characteristics of the main high-energy linear \ee \ colliders under consideration. Focusing on the newest such proposal, \CCCnospace, we demonstrated how the beam parameters can be modified to achieve around $40 \%$ higher instantaneous luminosity while maintaining the same level of beam-induced background. We also discussed potential further luminosity enhancements by adjusting the number of bunches per train and the train repetition rate, although detailed studies of the accelerator and detector design are necessary to assess the feasibility of such scenarios. Finally, we compared the luminosity spectra and background rates of \CCC with CLIC and ILC and concluded that ILC-250 and \CCCnospace-250 have the narrowest luminosity spectra, whereas CLIC and \CCCnospace-550 have the widest ones. Nevertheless, the expected number of background particles per bunch train for \CCC is lower than the respective number for CLIC and ILC at both center-of-mass energies, indicating that the pair-produced background at \CCC is manageable within the detector designs developed for these colliders. Our results suggest that \CCC is an attractive alternative for realizing a compact linear \ee \ collider that can reach the same or higher luminosities under similar or reduced background levels. Moreover, the developed methodology for luminosity optimization is quite generic, allowing it to be applied to other colliders, both linear and circular, beyond \CCCnospace.





\begin{acknowledgments}
The authors express their gratitude to Glen White, Wei-Hou Tan, Martin Breidenbach, and Lindsey Gray for their insightful discussions, which have significantly contributed to this study.
The work of the authors is supported by the US Department of Energy under contract DE–AC02–76SF00515.
\end{acknowledgments}

\newpage 

\bibliographystyle{atlasnote}

\bibliography{sample}

\end{document}